\documentclass{article}
\pdfoutput=1
\usepackage{jcappub} 
\usepackage{multirow}
\usepackage{xcolor}
\usepackage{graphicx}
\usepackage{hyperref}

\definecolor{darkblue}{RGB}{14,0,185}

\title{Gravitational wave background from primordial black holes in globular clusters}

\author[a,b]{Eleonora Vanzan,}
\emailAdd{eleonora.vanzan@phd.unipd.it}

\author[d,a]{Sarah Libanore,}
\emailAdd{libanore@bgu.ac.il}

\author[a,b]{Lorenzo Valbusa Dall'Armi,}
\emailAdd{lorenzo.valbusadallarmi@phd.unipd.it}

\author[a,b,c]{Nicola Bellomo,}
\emailAdd{nicola.bellomo@unipd.it}

\author[a,b,e,f]{Alvise Raccanelli}
\emailAdd{alvise.raccanelli.1@unipd.it}

\affiliation[a]{Dipartimento di Fisica e Astronomia G. Galilei, Università degli Studi di Padova, via Marzolo 8, I-35131 Padova, Italy.}
\affiliation[b]{INFN, Sezione di Padova, Via Marzolo 8, I-35131, Padova, Italy.}
\affiliation[c]{Texas Center for Cosmology and Astroparticle Physics, Weinberg Institute, Department of Physics, The University of Texas at Austin, Austin, TX 78712, USA.}
\affiliation[d]{Department of Physics, Ben-Gurion University of the Negev, Be'er Sheva 84105, Israel.}
\affiliation[e]{INAF - Osservatorio Astronomico di Padova, Vicolo dell'Osservatorio 5, I-35122 Padova, Italy.}
\affiliation[f]{Theoretical Physics Department, CERN, 1 Esplanade des Particules, 1211 Geneva 23, Switzerland.}

\begin{abstract}
{Primordial black holes still represent a viable candidate for a significant fraction, if not for the totality, of dark matter.
If these compact objects have masses of order tens of solar masses, their coalescence can be observed by current and future ground-based gravitational wave detectors.
Therefore, finding new gravitational wave signatures associated with this dark matter candidate can either lead to their detection or help constraining their abundance.
In this work we consider the phenomenology of primordial black holes in dense environments, in particular globular clusters.
We model the internal structure of globular clusters in a semi-analytical fashion, and we derive the expected merger rate.
We show that, if primordial black holes are present in globular clusters, their contribution to the GW background can be comparable to other well-known channels, such as early- and late-time binaries, thus enhancing the detectability prospects of primordial black holes and demonstrating that this contribution needs to be taken into account.}
\end{abstract}

\begin{document}

\maketitle

\section{Introduction}

Gravitational wave (GW) astronomy opened the door to the exploration of the Universe through gravitational radiation~\cite{abbott:firstligodetection, abbott:multimessenger}. 
With already~$\mathcal{O}(100)$ GW events detected from the LIGO-Virgo-KAGRA Collaboration~\cite{abbott:gwtc1, abbott:gwtc2, abbott:gwtc3}, we started constraining both astrophysical properties of the sources~\cite{abbott:O12eventsproperties, abbott:O3eventsproperties} and cosmological properties of the Universe~\cite{abbott:H0measurementI, abbott:H0measurementII, abbott:H0measurementIII}.
For instance, a single binary neutron star event with an electromagnetic counterpart has been remarkably effective in ruling out entire classes of modified gravity theories~\cite{creminelli:bnsconstraintsonmg, ezquiaga:bnsconstraintsonmg, baker:bnsconstraintsonmg}.
Given the great potential of the next observational runs, it is paramount to find novel ways to exploit this new probe for cosmological purposes, especially because GWs have the potential to answer some of the open questions in cosmology.

One very exciting possibility is related to shedding light on the nature of dark matter~\cite{bertone:gwprobesofdarkmatter}. 
So far, many viable candidates have been proposed, see, e.g.,~refs.~\cite{bertone:dmreviewI, bertone:dmreviewII}, including Weakly Interacting Massive Particles (WIMPs)~\cite{Jungman_1996, Bergstr_m_2000}, axions~\cite{Turner:1989vc, Carosi_2013, Marsh_2016}, sterile neutrinos~\cite{Kusenko_2009}, and Primordial Black Holes (PBHs).
Despite being initially proposed in the early '70s~\cite{zeldovich:pbhformation, hawking:pbhformation, carr:pbhformation, chapline:pbhformation}, PBHs received a newly found interest after the first LIGO detection~\cite{bird:pbhlatetimebinaries, clesse:pbhlatetimebinaries, sasaki:earlytimebinaries}. 
In contrast to ordinary astrophysical black holes (ABHs) created at the end of the stellar cycle of very massive stars, PBHs form (in the most standard scenario) in the early Universe, during the radiation-dominated era.
Moreover, since their formation mechanism is not related to any astrophysical process, they can in principle have any mass.

One guaranteed signature of this dark matter candidate is GW emission from dynamical processes involving PBHs, such as binaries' inspirals and mergers, see, e.g.,~refs.~\cite{carr:pbhconstraintsreview, sasaki:pbhconstraintsreview}.
In fact, current constraints on PBH abundance come from a variety of different probes, see, e.g.,~ref.~\cite{bagui:lisapbhreview,Caprini:2018mtu} for recent reviews, including GW experiments.
Depending on the nature of the process and the mass of PBHs, the characteristic GW emission can peak over a broad frequency range: from the nHz band probed by pulsar timing arrays~\cite{NANOGrav:2023gor, EPTA:2023fyk, Reardon:2023gzh, Xu:2023wog, InternationalPulsarTimingArray:2023mzf}, to the mHz-dHz band probed by future space-borne interferometers (e.g.,~LISA~\cite{amaroseoane:lisawhitepaper}, Big Bang Observer~\cite{phinney:bigbangobserverwhitepaper}, and DECIGO~\cite{kawamura:decigowhitepaperI, kawamura:decigowhitepaperII}), up to the Hz-kHz band targeted by current and future ground-based GW observatories (e.g.,~LIGO, Virgo, KAGRA, Einstein Telescope~\cite{punturo:einsteintelescopewhitepaper, maggiore:einsteintelescopewhitepaper}, and Cosmic Explorer~\cite{reitze:cosmicexplorerwhitepaper, evans:cosmicexplorerwhitepaper}).
Therefore, PBHs are expected to leave an imprint on top of the gravitational wave background (GWB) sourced by other known astrophysical sources, such as astrophysical black holes of any mass, neutron stars and white dwarfs.\footnote{In this work, we do not consider the so called cosmological GWB, sourced in the very early Universe, see e.g.,~\cite{Caprini:2018mtu} for a review. In that case, the process of PBH formation can further contribute to GW emission.}
It then becomes extremely timely and important to characterize PBH GW signatures, and possibly find novel techniques to isolate PBH contributions from the expected astrophysical ones.

Several authors already analyzed detectability prospects of PBH binaries~\cite{raccanelli:gwxlss, scelfo:gwxlssI, scelfo:gwxlssII, mukherjee:2021ags, mukherjee:2021itf, bosi:gwxlss, libanore:gwxlss} that formed either in the early Universe before matter-radiation equality~\cite{sasaki:earlytimebinaries, raidal:earlytimebinariesI, alihaimoud:earlytimebinaries}, or at late times in small dark matter halos~\cite{bird:pbhlatetimebinaries, clesse:pbhlatetimebinaries}.
However, limited attention has been payed to PBH phenomenology in dense astrophysical environments, such as stellar clusters~\cite{Kritos2020, Kou:2024gvp}.
Since it is widely recognized that compact object binaries in different kinds of cluster strongly contribute to GW emission due to enhanced interaction rates~\cite{Kou:2024gvp}, it is also reasonable to expect a strong PBHs contribution coming from the very same type of environments.
If so, current estimates of the GWB are missing a potentially significant contribution coming not only from PBH-PBH and PBH-ABH binaries, but also from GW bursts generated by compact objects flybys.
Several studies in recent years highlighted the role of ABHs in globular~\cite{Benacquista_2013, rodriguez_2015, fragione_2018}, nuclear~\cite{oleary_2009, Miller_2009, Zevin_2017, Neumayer_2020, Zevin_2021} and young~\cite{yc_2010, yc_2020, yc_2021, yc_2022, yc_2022_1} clusters.
Despite all these types of star clusters being gravitationally bound systems, they have substantially different formation histories and properties.
In particular, globular clusters (GCs) represent the most interesting candidate to study the impact of PBHs on dense environments, because of their early formation time and large central density.

This work aims to provide, for the first time, a comprehensive modeling of the GWB produced by PBH dynamical encounters in dense environments. 
We build a semi-analytical model that accounts for PBHs in GCs, and we derive from first principles all the relevant probability distribution functions that regulate gravitational interactions between different kinds of compact objects in the cluster.
Then we estimate the contribution to the GWB generated not only by binaries formed via direct capture and three-body processes, but also by hyperbolic encounters.
The former contribution peaks in the frequency band probed by the ground-based GW observatories, and we show that this binary formation channel is potentially detectable for next generation instruments.
Moreover, we show that this novel PBH GW signature can be more prominent than those of other more established channels, such as early- and late-time PBH binaries.
On the other hand, despite the large interaction rate, the hyperbolic encounters contribution peaking in the mHz band will not be detectable by future space-borne GW detectors.

The paper is organized as follows.
In section~\ref{sec:PBH} we describe PBH binary formation channels, and we start introducing a coherent model of GCs that incorporates PBHs.
The GC model is fully characterized in section~\ref{sec:globular_clusters}.
In section~\ref{sec:GWB} we compute the PBH contribution to the GWB, including different kinds of dynamical interaction in different frequency bands, and we compare it to the more established contributions, both primordial and astrophysical.
Finally, we conclude in section~\ref{sec:conclusions}.
Appendices contain additional information regarding typical timescales associated with the GC model (appendix~\ref{app:time_scales}), relative velocity distributions of compact objects inside GCs (appendix~\ref{app:relative_velocity_pdf}), hyperbolic encounters dynamics (appendix~\ref{app:hyperbolic_encounter_dynamics}) and probability distribution functions (appendix~\ref{app:hyperbolic_encounter_pdfs}).

\textbf{Fiducial Cosmology.}
In this work we assume a spatially flat $\Lambda$CDM cosmology~\cite{aghanim:cosmologicalparameters} with present-day Hubble expansion rate~$H_0 = 67 \, \mathrm{km/s/Mpc}$, present-day matter and radiation relative density~$\Omega_{\mathrm{m}0} = 0.307$ and~$\Omega_{\mathrm{r}0} = 9.061\times 10^{-5}$, respectively.


\section{Primordial black holes}
\label{sec:PBH}

In the most common scenarios, PBHs form before Big Bang Nucleosynthesis from the gravitational collapse of large density perturbations; see, e.g.,~\cite{sasaki:pbhconstraintsreview, carr:pbhconstraintsreview} for extensive reviews of different formation mechanisms.
However, regardless of their origin, only PBHs with masses larger than~$M_\mathrm{PBH} \gtrsim 10^{-18} \, M_{\odot}$ evaporate via Hawking radiation on a timescale larger than a Hubble time, and are considered as a viable dark matter candidate.
Their abundance is parameterized by the relative abundance parameter~$f_\mathrm{PBH} = \bar{\rho}_\mathrm{PBH} / \bar{\rho}_\mathrm{dm}$, where~$\bar{\rho}_\mathrm{PBH}$ and~$\bar{\rho}_\mathrm{dm}$ are the background PBH and dark matter energy densities.
The mass distribution of PBHs is described as~\cite{bellomo:pbhconstraints} 
\begin{equation}
    \frac{df_\mathrm{PBH}}{dM_\mathrm{PBH}} = f_\mathrm{PBH} \frac{d\Phi_\mathrm{PBH}}{dM_\mathrm{PBH}} \, ,
\label{eq:fpbhdm}
\end{equation}
where~$d\Phi_\mathrm{PBH}/dM_\mathrm{PBH}$ encodes the shape of the PBH mass distribution, and it is normalized to unity.
Different formation mechanisms typically produce different mass distributions, however in the following we focus exclusively on the so-called monochromatic case, widely used in the literature, where~$d\Phi_\mathrm{PBH}/dM_\mathrm{PBH} = \delta^D(M_\mathrm{PBH}-M^\star_\mathrm{PBH})$ is a Dirac delta centered at mass $M^\star_\mathrm{PBH}$; further discussion on this can be found in section~\ref{subsec:pbh_binary_signatures}.
Our results can be converted so to account for an extended mass function e.g.,~by following the approach in ref.~\cite{bellomo:pbhconstraints}.

While current abundance constraints appear to rule out PBHs as a significant constituent of dark matter~\cite{sasaki:pbhconstraintsreview, carr:pbhconstraintsreview} on some mass ranges, these upper bounds on~$f_\mathrm{PBH}$ suffer from many kinds of uncertainty.
In particular, constraints in the~$\mathcal{O}(10)\ M_\odot$ mass range have been showed to not be robust under changes in the PBH mass distribution, see, e.g.,~ref.~\cite{bellomo:pbhconstraints}, or under changes in the modeling of the observable, see, e.g.,~refs.~\cite{boschramon:pbhfeedbackI, boschramon:pbhfeedbackII, piga:pbhfeedback}.
Since in the next sections we are mainly interested in GW emission in the frequency band probed by current and future ground-based GW observatories, we assume as benchmark values for PBH masses~$M_{\rm PBH} = 1, 10, 100\, M_{\odot}$.
Because of the aforementioned uncertainties, we test different values for $f_\mathrm{PBH}$, including the extreme case~$f_\mathrm{PBH}=1$ in which PBHs constitute the totality of dark matter.
Finally, we assume PBHs to be spinless.

In this work we investigate three main PBH contributions to the GWB: early- and late-time binaries, in sections~\ref{subsec:early_time_pbh_binaries} and~\ref{subsec:late_time_pbh_binaries}, respectively, and the novel PBH contribution sourced in dense environments, which we introduce in section~\ref{subsec:gc_pbh_contribution}.
We discuss additional contributions that will be the subject of future work in section~\ref{sec:conclusions}.


\subsection{Early-time primordial black hole binaries}
\label{subsec:early_time_pbh_binaries}

Even during the radiation-dominated era, pairs of PBHs can form bound binary systems if they decouple from the background expansion~\cite{Nakamura:1997sm, Ioka:1998nz, sasaki:earlytimebinaries}.
Tidal forces generated by surrounding bodies and by the overall dark matter field (in the case PBHs do not constitute the totality of dark matter) avoid heads-on collisions and delay the merger, making the signal potentially observable today~\cite{alihaimoud:earlytimebinaries,  raidal:earlytimebinariesI, Raidal:2018bbj}. 
Following the formalism outlined in ref.~\cite{bosi:gwxlss}, we parameterize the merger rate of early PBH binaries for a monochromatic PBH mass function as
\begin{equation}
    R_{\rm EPBH}(z) = f_{\rm PBH}^{53/37} \mathcal{A}_m \left(\frac{t_0}{t(z)}\right)^{34/37}   \left(\frac{M_{\rm PBH}}{30\ M_\odot}\right)^{-32/37} \, ,
\label{eq:PBHrate_early}
\end{equation}
where~$t(z)$ is the cosmic time at the time of the merger, $t_0=t(0)$ is the age of the Universe, and~$\mathcal{A}_m$ is an amplitude parameter.
This formulation was originally derived in ref.~\cite{alihaimoud:earlytimebinaries}, based on the initial distribution of the orbital parameters of PBH pairs in the radiation-domination era.
While the dependencies on~$f_{\rm PBH},t(z)$ and~$M_{\rm PBH}$ are analytically derived, there are significant uncertainties on the amplitude $\mathcal{A}_m$.
This parameter depends on the properties of the environment at formation and during binary evolution across cosmic times~\cite{raidal:earlytimebinariesI, Ballesteros:2018swv, DeLuca:2020jug, Garriga:2019vqu, Hayasaki:2009ug}; its value is estimated from numerical simulations.
Refs.~\cite{Jedamzik:2020ypm,Jedamzik:2020omx} showed that~$\mathcal{A}_m \sim \mathcal{O}(10) \, {\rm Gpc^{-3}yr^{-1}}$ when only early and ABH binaries are considered.
In section~\ref{subsec:pbh_binary_signatures}, we consider the two cases~$\mathcal{A}_m = \{0.37,12\} \, {\rm Gpc^{-3}yr^{-1}}$ when~$f_{\rm PBH}=1$.
These values allow us to obtain a PBH merger rate (including early and late binaries, see next section) that does not exceed LVK constraints and is not in contrast with expectations on the merger rate of ABH binaries from stellar evolution models.


\subsection{Late-time primordial black hole binaries}
\label{subsec:late_time_pbh_binaries}

Another well-known PBH binary formation channel is given by direct capture processes in low-mass dark matter halos~\cite{bird:pbhlatetimebinaries, clesse:pbhlatetimebinaries}.
In this kind of environments, where relative velocities between halo objects are relatively low, unbounded PBHs can form binaries via the emission of GWs.
Also in this case, we follow ref.~\cite{bosi:gwxlss} and we parameterize the late-time binary merger rate as
\begin{equation}
    R_{\rm LPBH}(z) = f_{\rm PBH}^2 \int dt_{\rm d} p(t_{\rm d}) \int dM_{\rm h} \frac{dn_{\rm h}}{dM_{\rm h}} \mathcal{R}_{{\rm BF,h}}(M_{\rm h},z) \, ,
\label{eq:PBHrate_late}
\end{equation}
where~$t_{\rm d}$ is the time delay between the binary formation and merger, $p(t_{\rm d})$ is the time delay probability distribution function, $M_{\rm h}$ is the mass of the host DM halo, $dn_{\rm h}/dM_{\rm h}$ the halo mass function, and~$\mathcal{R}_{{\rm BF,h}}(M_{\rm h},z)$ is the binary formation rate per halo due to a direct capture process. 
This parametrization has been widely adopted in the literature, starting from refs.~\cite{bird:pbhlatetimebinaries,bellomo:pbhconstraints}, and it relies on the fact that the PBH number density inside an halo -- which enters squared in the two body process in  equation~\eqref{eq:PBHrate_late} -- can be expressed in terms of its radial density profile.
Finally, as detailed in section~\ref{subsubsec:DC}, the rate per halo~$\mathcal{R}_{\rm BF,h}$ depends on the relative velocity between the two PBHs.
Its value gets renormalized in the analysis such that the total rate late-binaries' merger rate is~$R_{\rm LPBH}(z) \simeq 2 \, {\rm Gpc}^{-3}{\rm yr}^{ -1}$ when~$f_{\rm PBH} = 1$.\footnote{This value is estimated in ref.~\cite{bird:pbhlatetimebinaries} by integrating a Navarro-Frenk-White density profile~\cite{Navarro_1996} for the DM halos, and a Tinker halo mass distribution~\cite{Tinker_2008}.}


\subsection{Primordial black holes in globular clusters}
\label{subsec:gc_pbh_contribution}

The dark matter content of GCs is largely debated.
On the observational side, the conclusions are extremely sensitive to the adopted priors and parametrizations -- see e.g.,~appendix C of ref.~\cite{garani:darkmatterinclusters} for a short summary of observational searches for dark matter in GCs.
On the simulation side, even though numerical simulations usually include dark matter at initial time~\cite{mashchenko:globularclustersimulations, baumgardt:globularclustersimulations, kippenhahn:globularclustersimulations, elbadry:globularclustersimulation,ardi:globularclustersimulation}, this component is either retained or tidally stripped by the host galaxy depending on the nature of dark matter and on which astrophysical prescriptions are included in the simulations.
The details of these processes are currently highly uncertain and, depending on the parameters of the simulation, the final outcome can be significantly different, see, e.g.,~refs.~\cite{Wirth:gcdm, Boldrini:gcdm, Carlberg:gcdm, Ma_2020, capela:pbhconstraints}.
Despite this underlying uncertainty, it is commonly accepted that, if dark matter is retained inside GCs, it is more likely to be found in their cores~\cite{heggie:dminglobularclusters, capela:pbhconstraints, hassani:dminglobularclusters}.
In particular, it has been shown that a dark matter component is expected in the core of GCs that formed at early times~$z \gtrsim 7$ inside dark matter minihalos~\cite{capela:pbhconstraints}, even after accounting for the action of the tidal field of the host galaxy~\cite{mashchenko:globularclustersimulations,mashchenko:globularclustersimulations_1}.

If PBHs account for at least part of the dark matter and remain bounded to the clusters across their evolution, the possibility of finding them in the core is even more likely, due to the process of mass segregation, described in the next section. In this case, we expect GC formation and evolution dynamics to be considerably influenced by the presence of PBHs in their core, contributing not only to shape the structure of such systems, but also to the total GW emission generated by the cluster components.

In the following, we model the PBH abundance in GCs by introducing the cluster abundance parameter
\begin{equation}
    f^\mathrm{cl}_\mathrm{PBH} = \frac{M^\mathrm{tot}_\mathrm{PBH}}{M_\mathrm{b}} \, ,
\label{eq:fpbhcl}
\end{equation}
where~$M_{\rm PBH}^{\rm tot}$ and~$M_\mathrm{b}$ are the total PBH and baryonic\footnote{As described in the next section, by baryonic mass we refer to stars and ABHs, while we neglect the presence of subdominant contributions, e.g.,~neutron stars, white dwarfs or interstellar gas.} mass of the cluster, respectively. 
Therefore, the total mass of a cluster is given by~$M_\mathrm{cl} = M_\mathrm{b}(1+f^\mathrm{cl}_\mathrm{PBH})$, and the number of PBHs per cluster for a monochromatic mass distribution reads as~$N_{\rm PBH} = M_{\rm b}f_{\rm PBH}^{\rm cl} / M_{\rm PBH}$.
In general, $f^\mathrm{cl}_\mathrm{PBH}$ is different from the relative abundance parameter~$f_\mathrm{PBH}$ introduced in equation~\eqref{eq:fpbhdm}. 
In the following sections we consider the benchmark cases for which~$f^\mathrm{cl}_\mathrm{PBH} = 0.1, 1$, and we argue about a possible link between this parameter and~$f_\mathrm{PBH}$ in section~\ref{subsec:globular_cluster_number_density}.


\section{Globular clusters}
\label{sec:globular_clusters}

Globular clusters are among the oldest structures in the Universe, they have an almost-spherical shape, large mass and a dense compact core.
In this work, we assume GCs are made of three different components, namely stars, ABHs, and PBHs, and we develop a semi-analytical model to account for their mass, velocity, and density distributions.
Hereafter, we refer to stars and ABHs as {\it baryonic mass components}, and we account for them in the~$M_{\rm b}$ parameter introduced in equation~\eqref{eq:fpbhcl}.
Our formalism can be generalized to other types of cluster, for instance young or nuclear clusters, by changing the characteristic mass, radius and redshift distributions.
While an accurate description of these environments can be obtained only via numerical simulations, our modeling represents a useful tool to understand the relevance of PBH GW signatures. 


\subsection{Stars and astrophysical black holes}
\label{subsec:ABH}

We assume that GCs at formation time are made only of small-intermediate $(M_\star \leq 8 \, M_\odot)$ and massive $(M_\star \geq 8 \, M_\odot)$ stars.
We model the star distribution at initial time as a Kroupa mass function~$dN_\star/dM_\star = A_\star M^{-\alpha}_\star$~\cite{kroupa:initialstarmassfunction}, where the normalization constant reads as
\begin{equation}\label{eq:Astar}
    A_\star = \frac{M^\mathrm{ini}_\mathrm{b}}{\displaystyle \int^{M^\mathrm{max}_\star}_{M_\star^\mathrm{min}} dM_\star M^{1-\alpha}_\star} \, ,
\end{equation}
$M^\mathrm{ini}_\mathrm{b}$ is the initial baryonic mass of the cluster, the scaling exponent is~$\alpha=2.3$, and the initial minimum and maximum star masses are taken to be~$M_\star^\mathrm{min}=0.5 \, M_\odot$ and~$M_\star^\mathrm{max}=40 \, M_\odot$, respectively.\footnote{
Other choices of the IMF are possible, however, this parametrization is extremely robust, as discussed, e.g.,~in ref.~\cite{Kroupa:2011aa} (see also ref.~\cite{Dickson2023} for a more in-depth discussion on GCs).
We checked that a~$2\sigma$ variability in the exponent of the power law, $\alpha = 2.3 \pm 0.36$~\cite{kroupa:2021,Kroupa:2011aa}, does not affect significantly the clusters' density and velocity profiles.}
Since observations seem to suggest that interstellar gas does not play a crucial role in GCs, we neglect its contribution.

The small-intermediate star population is characterized by a long lifetime and negligible mass loss during its evolution; hence, we assume that the low-mass tail of the mass distribution does not evolve in time and we do not account for small mass compact objects.
Conversely, stars in the massive population have shorter lifetimes and form BHs at the end of their stellar cycle.

We assume a negligible mass loss during the stellar lifetime and during the gravitational collapse that leads to the ABH formation, i.e.,~$M_\mathrm{ABH} \simeq M_\star$, and we expect a population of ABHs to form on a timescale of order~$\mathcal{O}(10)\ \mathrm{Myr}$ (see also appendix~\ref{app:time_scales}) in the mass range~$M_\mathrm{ABH}^\mathrm{min} = 8 \, M_\odot \leq M_\mathrm{ABH} \leq 40 \, M_\odot = M_\mathrm{ABH}^\mathrm{max}$.\footnote{
Stars with masses of order~$100 \, M_\odot$ can also form in GCs.
However, those stars are expected to suffer from a strong stellar mass loss before entering the BH formation stage~\cite{kippenhahn:globularclustersimulations}.
Hence, even choosing a larger maximum star mass, we still produce BHs in the same mass range we are considering here.
Given that the relative number of very massive stars is rather low because of the steepness of the star mass function, we choose~$M_\star^\mathrm{max}=40 \, M_\odot$, avoiding any modeling related to stellar mass loss.}
After all massive stars have disappeared, the ABH mass function can only evolve due to hierarchical mergers or cluster evaporation.
Regarding the former, simulations~\cite{mapelli_2021} suggest that the evolution of the mass function is weak in the mass range of interest, thus we neglect it in our parametrization.
On the other hand, evaporation can be prompted by large kick velocities acquired by compact objects during the supernova process.
We estimate this effect by introducing a retention fraction parameter~$f_{\rm ret}$ to rescale the initial baryonic mass in ABHs and account for such mass loss.
Typical values for the retention fraction are~$f_\mathrm{ret} \lesssim 1/2$, but exact values are still largely debated in the literature~\cite{pavlik:fret, morscher:fret_3bbinaryformationrate}. 
Following ref.~\cite{pavlik:fret} we define 
\begin{equation}
    f_{\rm ret} = \int_{0}^{v_{\rm esc}} dv_{\rm kick}\,p(v_{\rm kick}) \, ,
\end{equation}
where~$v_{\rm kick}$ is the natal kick velocity after the supernova event and~$v_{\rm esc}$ is the GC escape velocity. 
We model the natal kick probability distribution function as a Maxwell-Boltzmann distribution 
\begin{equation}
    p(v_{\rm kick}) = \frac{4 \pi v_{\rm kick}^2}{\left(2 \pi \sigma^2_\mathrm{kick}\right)^{3/2}} e^{-v_{\rm kick}^2 /2\sigma_{\rm kick}^2} \, ,
\end{equation}
and we verify that, for a typical velocity dispersion of~$\sigma_{\rm kick} \simeq 10 \, {\rm km/s}$~\cite{pavlik:fret}, GCs with baryonic mass~$M_\mathrm{b} \simeq 10^5 \, M_\odot$ have~$f_{\rm ret} \simeq 0.3$. 
Thus we set~$f_\mathrm{ret}=1/3$ as conservative fiducial value for ABHs in all GCs, independently on the cluster mass. 
Because of the mass loss due to natal kicks, the (final) baryonic mass of the cluster is lower than the initial one, i.e.,~$M_\mathrm{b} < M^\mathrm{ini}_\mathrm{b}$.
We assume that after ABHs with large natal kick velocities have left the cluster the mass functions of small-intermediate stars and of the remaining ABHs do not evolve further, therefore $M_{\rm b}$ can be effectively considered as a constant. 

In summary, star and ABH populations are characterized by a number of objects and a total mass that read as
\begin{equation}
    N_\star = A_\star \int^{M^\mathrm{min}_\mathrm{ABH}}_{M^\mathrm{min}_\star} dM_\star M^{-\alpha}_\star \, , \qquad M^\mathrm{tot}_\star = A_\star \int^{M^\mathrm{min}_\mathrm{ABH}}_{M^\mathrm{min}_\star} dM_\star M^{1-\alpha}_\star \, ,
\end{equation}
and
\begin{equation}
    N_\mathrm{ABH} = A_\star \int_{M^\mathrm{min}_\mathrm{ABH}}^{M^\mathrm{max}_\mathrm{ABH}} dM_\star M^{-\alpha}_\star f_\mathrm{ret} \, , \qquad M^\mathrm{tot}_\mathrm{ABH} = A_\star \int_{M^\mathrm{min}_\mathrm{ABH}}^{M^\mathrm{max}_\mathrm{ABH}} dM_\star M^{1-\alpha}_\star f_\mathrm{ret} \, ,
\end{equation}
respectively, explicitly defining the baryonic mass of the cluster~$M_\mathrm{b} = M_\star^\mathrm{tot} + M^\mathrm{tot}_\mathrm{ABH}$.


\subsection{Globular cluster internal structure}
\label{subsec:globular_cluster_modeling}

The evolution of the internal structure of a GC depends on many physical effects, for instance stellar evolution, $N$-body interactions, the effect of the tidal field of the hosting galaxy~\cite{heggie:globularclustersbook}.
Despite this intrinsic complexity, observational properties can be recovered under a modest set of assumptions that capture the main features of these systems~\cite{gunn:globularclusters, meylan:globularclusterreview}.

The spatial and velocity distribution of objects inside a cluster can be well represented by a distribution function~$f(v,r)$, where~$v$ is the velocity and~$r$ the distance from the cluster center.
The energy of an object in the cluster is defined as~$E(v,r)=\frac{1}{2}v^2-\Psi(r)$, where~$\Psi(r)$ is the effective gravitational potential, defined in such a way that $\Psi(r_{\rm t})=0$ at the truncation radius~$r_{\rm t}$, where objects are effectively stripped away from the cluster.
In this work, we consider an isotropic multi-mass King model~\cite{king:globularclusterdf, dacosta:multimasskingmodel}\footnote{
Generalizations of this model, for instance including anisotropies, are also employed in the literature, see, e.g.,~ref.~\cite{michie:anisotropicdf}; however, the King profile describes the cluster dynamics well enough for our purposes.} of the form
\begin{equation}
    f_j(E) = \frac{\rho_{0j}}{(2\pi\sigma^2_j)^{3/2}\mathcal{I}_{0j}} \left( e^{-E(v,r)/\sigma^2_j}-1 \right) \Theta_H \left(-E(v,r)\right) \, ,
\label{eq:king_phase_space_distribution}
\end{equation}
where~$j=\{\rm stars, \, ABH, \, PBH\}$ are the GC components, $\Theta_H(x)$ is the Heaviside Theta function and~$\sigma_j$ is a velocity scale, which approaches the value of the 1-dimensional velocity dispersion in the limit of very concentrated clusters, i.e.,~when~$\sqrt{\Psi}/\sigma_j \gg 1$.
The normalization constant $\mathcal{I}_{0j}$ is defined to recover the central density of each component, which is labelled as~$\rho_j(\Psi_0) = \rho_{0j}$. 
Velocity and density profiles are indirectly linked via the gravitational potential through
\begin{equation}
    \rho_j(\Psi) = 4\pi \int_0^{\sqrt{2\Psi}} dv\ v^2 f_j(E) = \frac{\rho_{0j}}{\mathcal{I}_{0j}} \left[ e^{\Psi/\sigma^2_j} {\rm erf}\left(\frac{\sqrt{\Psi}}{\sigma_j}\right) - \sqrt{\frac{4\Psi}{\pi\sigma^2_j}} \left( 1+\frac{2\Psi}{3\sigma^2_j} \right) \right] \, ,
\end{equation}
where the radial dependence on~$r$ is left implicit in~$\Psi$ and the integration upper bound is the escape velocity~$v_\mathrm{esc}(r)=\sqrt{2\Psi(r)}$.
The total density in the center of the cluster, $\rho_0 = \sum_j \rho_{0j}$, is determined by the Poisson equation
\begin{equation}
    \frac{d}{dr}\left( r^2\frac{d\Psi}{dr} \right) = -4\pi G r^2 \sum_j \rho_j(\Psi) \, ,
\end{equation}
where~$G$ is the Newton constant.
The velocity scale for each component~$\sigma_j$ is set by energy equipartition:\footnote{
More sophisticated models of the cluster internal structure account for the evolution in the density profile driven by interactions and energy exchange between different bodies. 
It has been shown that more massive bodies increase their velocities through these interactions, which in turn can lead to evaporation from the cluster and to the dynamical evolution of the overall structure across cosmic time. 
For this reason, clusters with a more complex dynamical model never reach the condition of energy equipartition~\cite{miocchi:energyequipartition, khalisi:energyequipartition, trenti:energyequipartition, torniamenti:energyequipartition}.
A careful modeling of these processes would require numerical simulations and goes beyond the scope of this work.}
because of two-body gravitational interactions, energy is redistributed between objects in the cluster, tending towards an equilibrium state~\cite{chandrasekhar:dynamicalfriction}. 
Under the (partial) equipartition condition, the velocities of two classes of objects with masses~$m_i$ and~$m_j$ scale as~$\sigma_i m^\delta_i = \sigma_j m^\delta_j$, where~$\delta\in [0,1/2]$, and the case~$\delta=1/2$ corresponds to a perfect energy equipartition scenario.

We use the public code \textsc{limepy}\footnote{\url{https://github.com/mgieles/limepy}}~\cite{gieles:limepy} to compute density and velocity profiles of GCs. 
In~\textsc{limepy}, we are required to specify the concentration parameter~$\Psi_0/\sigma^2$, where~$\sigma=(m_j/\bar{m})^\delta \sigma_j$ is the velocity scale corresponding to the central density weighted mean mass~$\bar{m}=\sum_j m_j \rho_{0j}/\rho_{0}$, and the half-mass radius~$r_{\rm h}$, i.e.,~the radius that encloses half of the mass of the cluster.
We consider a benchmark value of~$\Psi_0/\sigma^2 = 10$ for the concentration parameter~\cite{gunn:globularclusters, oleary:hardnessratio}, and of
\begin{equation}
    r_{\rm h} = 3 \left( \frac{M_\mathrm{cl}}{{10^5 \, M_\odot}} \right)^{1/3} \, \mathrm{pc} \, ,
\end{equation}
for the half-mass radius.
These values are compatible with existing fit to observations, see, e.g.,~ref.~\cite{henaultbrunet:milkywaycluster}.

\begin{figure}[ht]
    \centerline{
    \includegraphics[width=\columnwidth]{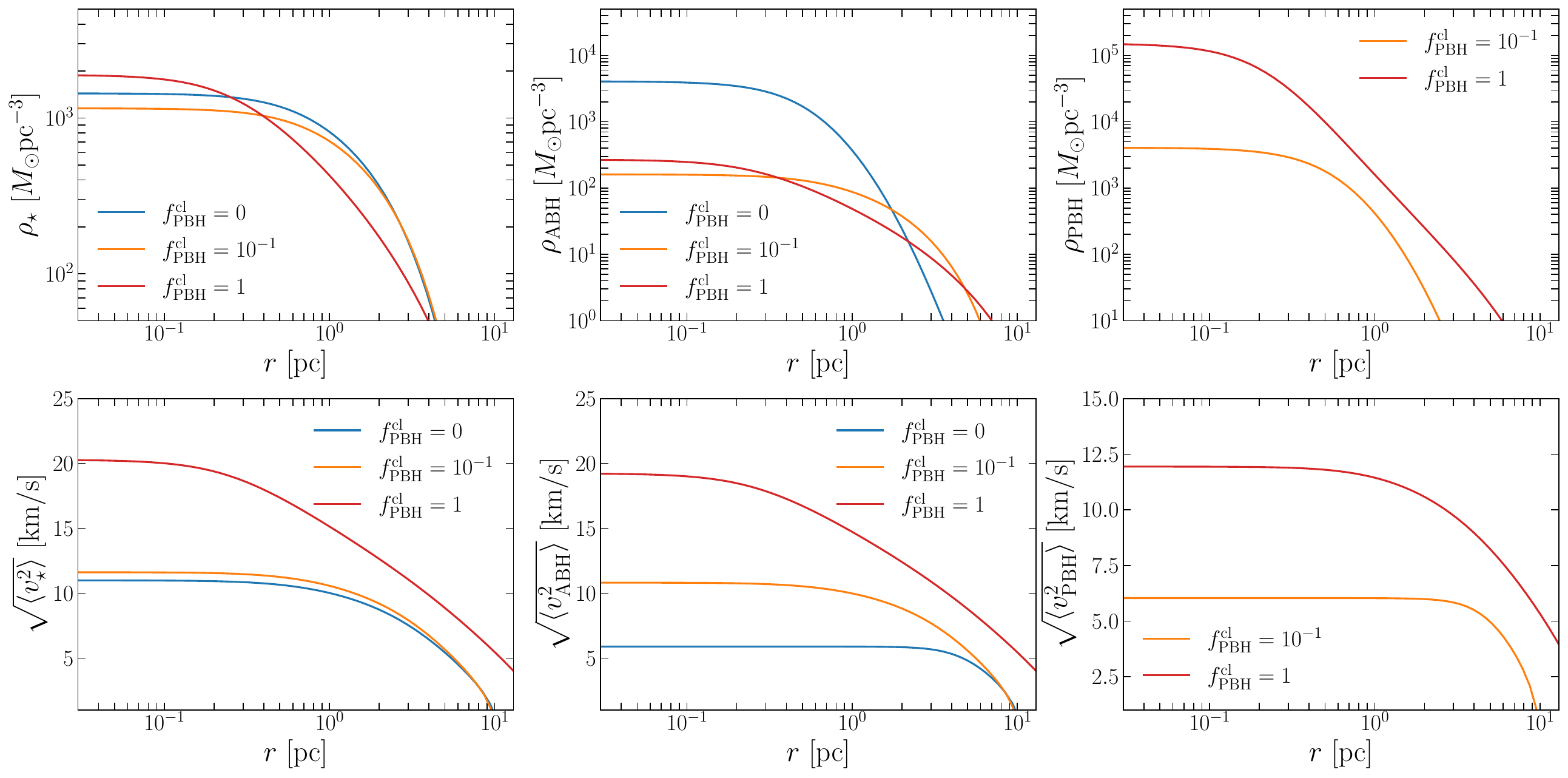}}
    \caption{Star (\textit{left panels}), ABH (\textit{central panels}) and PBH (\textit{right panels}) density (\textit{top panels}) and average squared velocity (\textit{bottom panels}) radial profiles for different value of the PBH cluster abundance parameter~$f^\mathrm{cl}_\mathrm{PBH}$.
    We show profiles for clusters with~$M_\mathrm{b} = 10^5 \, M_\odot$ and~$M_\mathrm{PBH}=100 \, M_\odot$.}
    \label{fig:cluster_profiles}
\end{figure}

As a result of gravitational interactions, more massive objects experience a dynamical friction effect, migrate towards the core of the GC, and become effectively segregated.
The typical timescale of the equipartition process is of the order of the half-mass relaxation time~\cite{spitzer:relaxationtime, binney:galacticdynamics},
\begin{equation}
    t_\mathrm{rel} = 0.17 \frac{N}{\log(\lambda N)} \sqrt{\frac{r^3_h}{GM_\mathrm{cl}}} \, ,
\label{eq:trelax}
\end{equation}
where~$N$ is the total number of objects in the cluster, and~$\lambda=0.11$ is a coefficient calibrated on numerical simulations~\cite{giersz:lambdacoefficient}.
We show in figure~\ref{fig:cluster_profiles} the stable configuration reached by clusters with different PBH abundance for~$M_\mathrm{PBH}=100\ M_\odot$ after a typical timescale of order~$t_\mathrm{rel} (M^\mathrm{ini}_\mathrm{b}=10^5\ M_\odot) \simeq 300\ \mathrm{Myr}$.
As we see from the bottom panels, increasing the PBH abundance consistently boosts the typical average squared velocity
\begin{equation}
    \begin{aligned}
        \left\langle v^2 \right\rangle_j &= \dfrac{\displaystyle 4\pi \int_0^{\sqrt{2\Psi}} dv\ v^4 f_j(E)}{\rho_j(\Psi)} \\
        &= 3\sigma^2_j  \dfrac{e^{\Psi/\sigma_j^2} {\rm erf} \left( \sqrt{\Psi}/\sigma_j \right) - \sqrt{4\Psi/\pi\sigma_j^2} \left( 1 + 2\Psi/3\sigma_j^2 + 4\Psi^2/15\sigma_j^4 \right)}{ e^{\Psi / \sigma^2_j} {\rm erf} \left( \sqrt{\Psi}/\sigma_j\right) - \sqrt{4\Psi/\pi\sigma_j^2} \left( 1 +  2\Psi/3\sigma_j^2 \right)} \, .
    \end{aligned}
\end{equation}
of each component, which more massive components (PBHs in the case displayed here) having a smaller average velocity because of energy equipartition. 
Density profiles are affected by the presence of massive PBHs in a non-trivial way: on the one hand, objects tend to be more concentrated because of the dense core formed by PBHs, on the other the increased average velocity tends to dilute objects lighter than PBH.
While these two effects appear to average out in the case of stars, we observe that the latter tend to dominate in the case of ABHs, which present a lower core density than in the~$f^\mathrm{cl}_\mathrm{PBH}=0$ case.
Historically, the size of the GC core has been described by the King radius~$r^2_0=9\sigma^2/(4\pi G \rho_0)$.
In our multi-mass King model we use the King radii~$r_{0j} = (\bar{m}/m_j)^\delta r_0$ as a measure of each component ``core'', where mass densities and velocities can be considered spatially uniform.
Outside this internal regions, density and velocity profiles rapidly decay, as can be seen in figure~\ref{fig:cluster_profiles}.
In the following section we will refer to these King radii as ``segregation radii''.


\subsection{Globular cluster number density}
\label{subsec:globular_cluster_number_density}

Assuming that GC mass distribution scales as~$(M^\mathrm{ini}_\mathrm{b})^{-2}$~\cite{O_Leary_2007, McLaughlin_2008, larsen_2008, elbadry:globularclustersimulation}, we have that
\begin{equation}
    \frac{dn_\mathrm{cl}}{dM^\mathrm{ini}_\mathrm{b}} = \frac{ m_\mathrm{b,tot}(z)}{\left(M^\mathrm{ini}_\mathrm{b}\right)^2 \log(M^\mathrm{ini}_\mathrm{b,max}/M^\mathrm{ini}_\mathrm{b,min})} \, ,
\end{equation}
where we consider~$M^\mathrm{ini}_{\rm b,min} = 5 \times 10^{4} \, M_\odot$ and~$M^\mathrm{ini}_{\rm b, max}=10^7 \, M_\odot$ as the minimum and maximum initial baryonic mass of a cluster, and~$m_\mathrm{b,tot}(z)$ is the total baryonic mass density in GCs at redshift~$z$.
The latter is defined as~$\displaystyle m_\mathrm{b,tot}(z) = \int^{t(z)}_0 dt\ \dot{m}_\mathrm{b}$, where the GC formation density rate is fitted to numerical simulations as~\cite{elbadry:globularclustersimulation, Rodriguez2018}
\begin{equation}
    \dot{m}_\mathrm{b}(z) = \frac{a}{\sigma \sqrt{2\pi}} \exp\left[-(z-\mu)^2/(2\sigma^2)\right] \left[ 1 + {\rm erf}\left( \frac{\alpha (z-\mu)}{\sqrt{2} \sigma} \right) \right] \, ,
\end{equation}
and the fitting parameters read as~$\{ \sigma, \mu, \alpha, a \} = \{ 2.7, 2.7, 1.5, 9.0\times 10^5 \, M_\odot \, \mathrm{Gpc^{-3} \, Myr^{-1}} \}$.
We find $n_\mathrm{cl}(z=0) \simeq 3 \ {\rm Mpc}^{-3}$, in agreement with current literature~\cite{PortegiesZwart:1999nm, Antonini_2020, Boylan-Kolchin:2017iji}.
The implementation of a time-evolving cluster mass function is left for future work.
Notice that the cluster mass function is expressed in terms of the initial mass~$M_{\rm b}^{\rm ini}$, which differs from the baryonic mass~$M_{\rm b}$ by the amount of ABHs that have been ejected from the cluster, as described in section~\ref{subsec:ABH}.

In section~\ref{sec:PBH} we introduced two abundance parameters to describe PBHs in the Universe,~$f_\mathrm{PBH}$, and in clusters,~$f^\mathrm{cl}_\mathrm{PBH}$.
While on the one hand we already showed in sections~\ref{subsec:early_time_pbh_binaries} and~\ref{subsec:late_time_pbh_binaries} that GW emission from early- and late-time binaries scales with powers of~$f_\mathrm{PBH}$, on the other hand we expect PBHs contribution from clusters to depend on~$f^\mathrm{cl}_\mathrm{PBH}$.
If we compare the total energy density in PBH ($\bar{\rho}_\mathrm{PBH}=f_\mathrm{PBH} \bar{\rho}_\mathrm{dm}$) to the total PBH energy density in GCs, which reads as 
\begin{equation}
    \begin{aligned}
        \rho^\mathrm{cl}_\mathrm{PBH}(z) &= \int dM^\mathrm{ini}_\mathrm{b} \frac{dn_\mathrm{cl}}{dM^\mathrm{ini}_\mathrm{b}} f^\mathrm{cl}_\mathrm{PBH} M_\mathrm{b} \\
        &= \left[1 - (1-f_\mathrm{ret}) \frac{\int_{M^\mathrm{min}_\mathrm{ABH}}^{M^\mathrm{max}_\mathrm{ABH}} dM_\star M_\star^{1-\alpha}}{\int_{M^\mathrm{min}_\star}^{M^\mathrm{max}_\star} dM_\star M_\star^{1-\alpha}} \right] f^\mathrm{cl}_\mathrm{PBH} m_\mathrm{b,tot}(z) \, ,
    \end{aligned}
\end{equation}
we observe that, for instance, at redshift~$z=0$, we can have the extreme case where GCs have equal parts of PBHs and baryons (i.e.,~$f^\mathrm{cl}_\mathrm{PBH}=1$), while having PBH relative abundance parameter as low as~$f_\mathrm{PBH} \simeq 2\times 10^{-5}$.
Even if this example is clearly extreme, it showcases that there are scenarios where PBH GW emission in certain channels is heavily suppressed while GW emission from dense environments is not.


\section{Primordial black hole gravitational wave background}
\label{sec:GWB}

In full generality, the total GWB in any frequency band can be written as the sum of independent GWBs generated by different sources and/or processes, i.e.,~$\Omega_\mathrm{GWB} = \sum_{\beta,i,j} \Omega^{(\beta)}_{ij}$, where~$\beta$ labels the different the process, while~$i,j$ indicate the sources (either ABH or PBH).
Not all combinations of dynamical processes~$(\beta)$ and sources~$(i,j)$ are necessarily possible; in the following, we only discuss the ones that are relevant for our work.

The individual backgrounds are given by~\cite{phinney:gwenergyspectrum, regimbau:astrophysicalsources}
\begin{equation}
    \Omega^{(\beta)}_{ij} = \frac{f_{\rm obs}}{\rho_{0c} c^2}\int \frac{dz \, d\boldsymbol{\theta} \, p(\boldsymbol{\theta},z)}{(1+z)H(z)}  \mathcal{R}_{ij}^{(\beta)} (\boldsymbol{\theta}, z) \frac{dE^{(\beta)}_{ij}}{df} (\boldsymbol{\theta}, z, f_{\rm obs}) \, ,
\label{eq:omega_sgwb}
\end{equation}
where~$f_{\rm obs} = f/(1+z)$ and~$f$ are the observed and emitted GW frequency, respectively, $z$ is the GW emission redshift, $\rho_{0c}$ is the present-day critical density, $c$ is the speed of light, $H(z)$ the Hubble expansion rate.
We label with $\boldsymbol{\theta}$ the set of parameters that describe both the process and the sources, and we introduce~$p(\boldsymbol{\theta},z)$ to characterize their probability distribution functions.
These are used to compute the dynamical process density rate $\mathcal{R}_{ij}^{(\beta)}$, and the GW energy spectrum~$dE^{(\beta)}_{ij}/df$.
The latter is commonly written as~\cite{maggiore:gwenergyspectrum}
\begin{equation}
    \frac{dE^{(\beta)}_{ij}}{df} = \frac{2\pi^2 D^2_L c^3 f^2_{\rm obs}}{G (1+z)^2}  \left| h^{(\beta)}_{ij} \right|^2 \, ,
\end{equation}
where~$D_L$ is the luminosity distance and~$ h^{(\beta)}_{ij}(\boldsymbol{\theta}, z, f_{\rm obs})$ is the emitted GW strain.

The dimensionality of the parameter space in equation~\eqref{eq:omega_sgwb} rapidly increases when accurate descriptions of dynamical processes are actually implemented, as in this work.
Apart from the emission redshift~$z$, we are interested in integrating over the GC initial baryonic masses~$M^\mathrm{ini}_\mathrm{b}$, the masses of the compact objects~$M_i,M_j$, and any other astrophysical parameter that describes the dynamical process at hand.
In order to make the problem numerically tractable, we choose to treat ABH and PBH mass distributions on equal grounds, i.e.,~as monochromatic mass functions.
While for PBHs this is already the case by assumption, we compute an effective ABH mass~$M_\mathrm{ABH}^\mathrm{eff}$ by requiring that the GWB produced by such monochromatic ABH mass distribution matches the true GWB produced by the real ABH mass distribution in the GC core.
In practice, the ABH effective mass for each process and combination of sources (in which at least one of them is an ABH) is computed by solving, either analytically or numerically, the equation
\begin{equation}
\label{eq:MeffABH}
    \Omega^{(\beta)}_{ij}\left( \frac{dn_\mathrm{ABH}}{dM_\mathrm{ABH}} = \frac{\rho_{0\rm ABH}}{M_\mathrm{ABH}^\mathrm{eff}} \delta^D(M_\mathrm{ABH} - M_\mathrm{ABH}^\mathrm{eff}) \right) = \Omega^{(\beta)}_{ij}\left( \frac{dn_\mathrm{ABH}}{dM_\mathrm{ABH}} = \frac{(2-\alpha) M^{-\alpha}_\mathrm{ABH} \rho_{0\rm ABH}}{(M^\mathrm{max}_\mathrm{ABH})^{2-\alpha} - (M^\mathrm{min}_\mathrm{ABH})^{2-\alpha}} \right) \, ,
\end{equation}
where on the right hand side we adopted the Kroupa mass function defined in section~\ref{subsec:globular_cluster_modeling}.

Equation~\eqref{eq:omega_sgwb} accounts for all GW emissions produced by the~$(i,j)$ objects in the process~$(\beta)$; depending on the signal-to-noise ratios, these will be either resolved events or part of the background.
Since the aim of this work is not to characterize detection prospects of the different GWB presented in this work, we do not distinguish between sources that are effectively resolved or not by a specific GW detector network. Instead, all the GWBs reported hereafter refer to the total contribution generated by all the sources in any given processes.
We leave accurate detectability forecasts for future works.


\subsection{Early- and late-time primordial black hole binaries}
\label{subsec:early_late_time_binaries_gwb}

Following ref.~\cite{bosi:gwxlss}, we compute the GWB generated by early- and late-time binaries using the external modules of~\texttt{CLASS\_GWB}~\cite{bellomo:classgwb}.
The external modules create a catalog of GW binaries by sampling the properties of the GW sources from the characteristic probability distribution functions of the parameters that describe the merging event.
For instance, in the case of early- and late-time PBH binaries, we sampled the merging redshift starting from the merger rates in equations~\eqref{eq:PBHrate_early} and~\eqref{eq:PBHrate_late}, their masses, spins, sky localization, inclination angle and polarization angle.
The GWB is then computed by summing over the contribution of all the sources in the catalog.
More detail on this procedure can be found in ref.~\cite{bellomo:classgwb}.


\subsection{Astrophysical black hole binaries}
\label{subsec:astrophysical_blackhole_binaries}

Another useful term of comparison for our results is given by the GWB generated by a generic population of astrophysical BHs with properties compatible with data from the latest LIGO-Virgo-KAGRA collaboration results~\cite{abbott:O12eventsproperties, abbott:O3eventsproperties}.
In this case, we do not specify the process that originated each binary, even if there are already some attempt to characterize different astrophysical formation channels~\cite{Zevin_2017}.
Also in this case, the GWB sourced by this population is computed using the external modules of~\texttt{CLASS\_GWB} following the model and the procedure described in ref.~\cite{bellomo:classgwb}.


\subsection{Dense environment binaries}
\label{subsec:globular_cluster_binaries}

The formation of binary BHs in dense stellar environments is driven by several different physical mechanisms, see, e.g.,~ref.~\cite{baibhav:binaryformationchannels} and references therein.
The dominant formation channel is expected to be the two-body direct capture ($\beta={\rm DC}$ hereafter) process~\cite{quinlan:directcapture, mouri:directcapture}, where pairs of unbounded compact objects end up in a bound system due to GW emission during their flyby.
However, other dynamical interactions are also possible; in this work, we account also for three-body interactions ($\beta = 3{\rm B}$)~\cite{baibhav:binaryformationchannels}, where the formation of the bound system is catalyzed by gravitational interaction with a third body. Other processes such as multiple encounters, hardening of the binary, disruption and exchanges of binary components with objects in the environment, are expected to provide a subdominant contribution~\cite{baibhav:binaryformationchannels}, thus we neglect them in our analysis. 

For each process and source pair, we define the interaction rate per compact object mass per cluster as 
\begin{equation}
\label{eq:Gamma_ij}
    \Gamma^{(\beta)}_{ij} = \left( 1-\frac{1}{2}\delta^K_{ij} \right) \frac{dn_i}{dM_i} \frac{dn_j}{dM_j} V_{\rm seg} \sigma^{(\beta)} v_{\rm rel} \, ,
\end{equation}
where the Kronecker delta~$\delta^K_{ij}$ avoids double counting for identical sources, i.e.,~when $i=j$. $V_{\rm seg} = {\rm min}[r_{0i},r_{0j}]$ is the volume defined by the smallest segregation radius, inside which number densities~$dn_{i,j}/dM_{i,j}$ are spatially uniform,~$v_{\rm rel}$ is the value of the relative velocity between the objects~$i$ and~$j$ and~$\sigma^{(\beta)}$ is the cross section of the dynamical process.
In the case of three body encounters, we estimate~$\sigma^(\beta)$ by marginalizing over the properties of the third body that catalyze the interaction.
Therefore, the total interaction rate inside each cluster is given by
\begin{equation}
\label{eq:Gamma_ij_tot}
    \Gamma_{ij,\mathrm{tot}}^{(\beta)} = \left( 1-\frac{1}{2}\delta^K_{ij} \right) \int dM_i dM_j \frac{dn_i}{dM_i} \frac{dn_j}{dM_j} V_{\rm seg} \left\langle \sigma^{(\beta)} v_{\rm rel} \right\rangle \, ,
\end{equation}
where we sum over the masses of the compact objects and we average the velocity-weighted cross section over the process-dependent parameters described by the probability distribution functions introduced in equation~\eqref{eq:omega_sgwb}.
In the previous equation, the quantities~$V_{\rm seg}, \, v_{\rm rel}, \, dn_{i,j}/dM_{i,j}$ depend on~$M_{\rm b}^{\rm ini}$, and they are estimated by relying on the \textsc{limepy}~\cite{gieles:limepy} public code, as described in section~\ref{subsec:globular_cluster_modeling}.


\subsubsection{Direct capture}
\label{subsubsec:DC}

The cross section in the direct capture case is~\cite{quinlan:directcapture, mouri:directcapture} 
\begin{equation}
\label{eq:bf_dc_cs}
    \sigma^\mathrm{(DC)} = 2\pi \left(\frac{85\pi}{6\sqrt{2}}\right)^{2/7} \frac{G^2 M^{12/7} \mu^{2/7}}{c^{10/7} v_{\rm rel}^{18/7}} \, ,
\end{equation}
where the total and reduced masses read as~$M = M_i+M_j$ and~$\mu = M_iM_j/M $, respectively.
In this case the velocity-weighted cross section has to account only for the relative velocity probability distribution function, i.e.,
\begin{equation}
    \left\langle \sigma^{(\mathrm{DC})} v_{\rm rel} \right\rangle = \int dv_\mathrm{rel} p(v_\mathrm{rel}) \sigma^{(\mathrm{DC})} v_{\rm rel} \, .
\end{equation}
We provide an analytical estimate of~$p(v_\mathrm{rel})$ in appendix~\ref{app:relative_velocity_pdf}.


\subsubsection{Three-body encounters}
\label{subsubsec:3B}

Three-body encounters can result in the formation of bounded binary systems, however only hard binaries are expected to survive~\cite{heggie:binarysurvival}.
Following the current literature, we consider only binaries with hardness ratio~$\eta$, i.e.,~the ratio between the average potential and kinetic energy of the environment~\cite{ivanova:hardnessratio, oleary:hardnessratio}, that is larger than the typical minimum value~$\eta_\mathrm{min} \simeq 5$~\cite{morscher:fret_3bbinaryformationrate}. 
Adapting the results of refs.~\cite{ivanova:hardnessratio, morscher:fret_3bbinaryformationrate, rodriguez:3bbinaryformationrate} to our formalism, we define the average velocity-weighted cross section for three-body encounters as
\begin{equation}
    \begin{aligned}
        \left\langle \sigma^{(\mathrm{3B})} v_{\rm rel} \right\rangle &= \sqrt{2} \pi^2 \frac{ G^5 (M_i+M_j)^5 }{\left\langle v_\mathrm{rel}\right\rangle^9} \frac{1+2\eta_{\rm min}}{\eta_{\min}^{5.5}} \\
        &\qquad\qquad \sum_{k=\mathrm{ABH,PBH}} s^{-1}_{ij,k} \int dM_k \frac{dn_k}{dM_k} \frac{V_{k,\mathrm{seg}}}{V_\mathrm{seg}} \left[ 1 +2\eta_{\rm min} \left(\frac{M_i+M_j+M_k}{M_i+M_j}\right) \right],
    \end{aligned}
\end{equation}
where~$\left\langle v_\mathrm{rel} \right\rangle = 2\sqrt{2/\pi}\sigma_\mathrm{rel}$ and~$\sigma_{\rm rel}^2=\sigma_i^2+\sigma_j^2$ are the average relative velocity and the relative velocity dispersion between bodies~$i$ and~$j$, respectively, $s_{ij,k}$ is a numerical factor that avoids double counting,\footnote{
We have~$s_{ij,k}=3$ if~$i=j=k$, $s_{ij,k}=2$ if~$i\neq j$ and~$s_{ij,k}=1$ if~$i=j\neq k$.}
and the three-body segregation volume is defined as~$V_{k, \rm seg} = {\rm min}[r_{0i}, r_{0j}, r_{0k}]$.


\subsubsection{Total binary contributions from globular clusters}
\label{subsec:total_binary_contribution_from_gc}

Independently on the formation channel, all binaries emit the same kind of GW signal during their inspiral, merger and ringdown phases.
In this work we use the templates provided in refs.~\cite{ajith:gwwaveformI, ajith:gwwaveformII, ajith:gwwaveformIII} to describe the GW strain~$h^\mathrm{(DC,3B)}_{ij}$ across those three phases of the coalescence process.
Each phase has its own unique dependency on the compact object masses; however, since in the frequency band of interest most of the emission is concentrated during the inspiral phase, we use the inspiralling mass dependence to compute the ABH effective mass.
Therefore, for this purposes, 
\begin{equation}
    \left| h^\mathrm{(DC,3B)}_{ij} \right|^2 \propto \mu M^{2/3}.
\end{equation}

\begin{figure}[ht]
    \centerline{
    \includegraphics[width=\columnwidth]{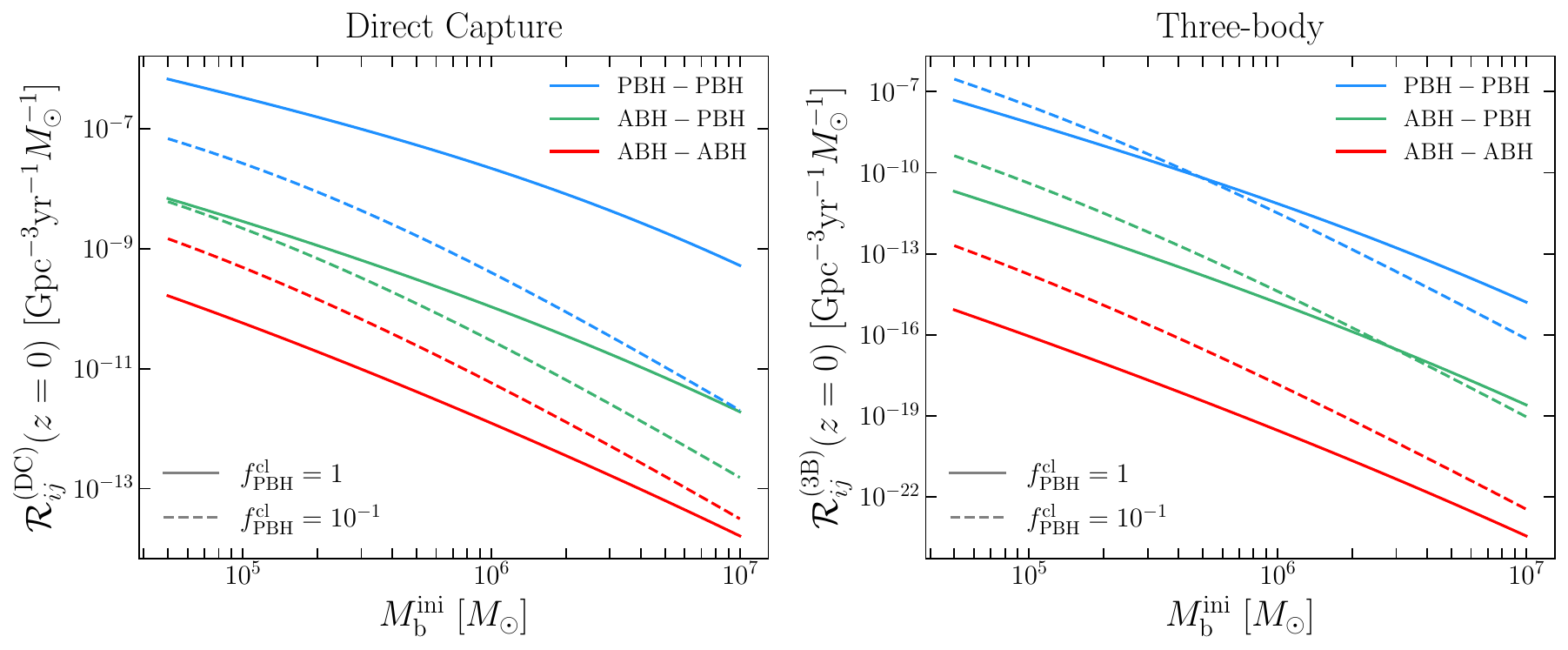}}
    \caption{Binary formation rate density at redshift~$z=0$ for direct capture (\textit{left panel}) and three-body (\textit{right panel}) processes. 
    In each panel we show ABH-ABH (\textit{blue lines}), ABH-PBH (\textit{green lines}) and PBH-PBH (\textit{red lines}) binaries, both for~$f_{\rm PBH}^{\rm cl}=1$ (\textit{solid lines}) and~$f_{\rm PBH}^{\rm cl}=10^{-1}$ (\textit{dashed lines}).
    The PBH mass is~$M_{\rm PBH}=100 \, M_{\odot}$ in all cases reported.}
    \label{fig:binary_formation_rate_densities}
\end{figure}

In figure~\ref{fig:binary_formation_rate_densities} we show the local, i.e.,~at redshift~$z=0$, binary formation rate density per cluster
\begin{equation}
    \mathcal{R}^{(\beta)}_{ij} (z, M^\mathrm{ini}_\mathrm{b}) = \frac{dn_{\rm cl}}{dM^\mathrm{ini}_{\rm b}} \Gamma_{ij,\rm tot}^{(\beta)}(M_{\rm b}^{\rm ini}) \, ,
\end{equation}
for the direct capture and three-body channels.
As showed in section~\ref{subsec:globular_cluster_modeling}, the addition of a PBH component induces a non-trivial change in the GC structure, both in terms of core densities and velocities, which translates into a non-trivial behaviour in the binary formation rate density.
First of all, we note the presence of a hierarchy between the binary formation rates per cluster: since PBHs form dense cores and dilute lighter ABHs, we have that the binary formation rates of the former population is larger than that of the latter one in both formation channels.
Due to (the disappearance of) this dilution effect, the binary formation rate of ABH-ABH systems increases when PBH abundance in GC decreases, and vice versa, consistently with the picture portrayed above.
Moreover, the direct capture channel appears to be the dominant contribution to the total binary formation rate in almost all cases considered, except for GCs with low mass and low PBH abundance.
While for direct capture the binary formation rate per cluster decreases when~$f^\mathrm{cl}_\mathrm{PBH}$ does so, in the case of three-body processes we observe the opposite behaviour for low mass GCs.
This seemingly counterintuitive effect is due to the fact that a lower PBH density is effectively compensated by a lower relative velocity dispersion of PBHs due to a shallower GC total potential well.

After they form, binaries merge in a characteristic time, which depends on the initial conditions of the compact object system.
Since those initial conditions are stochastic, each formation channel is characterized by its own time delay probability distribution function~$p(t_{\rm d})$, see appendix~\ref{app:time_scales}.
Therefore, the merger rate density is defined as
\begin{equation}
    R_{ij}^{(\beta)}(z) =  \int dt_{\rm d} p(t_{\rm d}) \int dM^\mathrm{ini}_{\rm b} \mathcal{R}^{(\beta)}_{ij} (z_\mathrm{bf}, M^\mathrm{ini}_\mathrm{b}) \, ,
\end{equation}
where the binary formation redshift~$z_\mathrm{bf}$ is defined implicitly for each time delay by~$z = z_{\rm bf}+t_{\rm d}$, and 
\begin{equation}
    t_{\rm d} = - \int_{z_\mathrm{bf}}^{z} \frac{dy}{(1+y)H(y)} \, .
\end{equation}
As we show in appendix~\ref{app:time_scales}, time delays for our processes of interest are negligible compared to the other timescales of the dynamical process, therefore we approximate~$p(t_{\rm d}) \simeq \delta^D(t_{\rm d})$.

In table~\ref{tab:merger_rate_densities}, we present  the typical values of the local, i.e., at redshift~$z=0$, merger rate densities for our benchmark models.

\begin{table}[ht]
    \centering
    \begin{tabular}{|ll|lll|}
    \hline
    \multicolumn{2}{|l|}{$R_{ij}^{\rm (DC+3B)}(z=0) \ [{\rm Gpc}^{-3} \, {\rm yr}^{-1}]$} & \multicolumn{1}{l}{ABH-ABH} & \multicolumn{1}{l}{ABH-PBH} & PBH-PBH \\ 
    \hline
    \hline
    \multicolumn{1}{|l|}{\multirow{3}{*}{$f^\mathrm{cl}_\mathrm{PBH}=1$}} & $M_\mathrm{PBH} = 1\ M_\odot$ & $2.8 \times 10^{-3}$ & $1.2 \times 10^{-3}$ & $4.4 \times 10^{-3}$ \\
    \multicolumn{1}{|l|}{} & $M_\mathrm{PBH} = 10\ M_\odot$ & $5.7 \times 10^{-3}$ & $2.4 \times 10^{-2}$ & $4.5\times 10^{-2}$ \\
    \multicolumn{1}{|l|}{} & $M_\mathrm{PBH} = 100\ M_\odot$ & $1.3 \times 10^{-5}$ & $7.8 \times 10^{-4}$ & $1.2\times 10^{-1}$ \\
    \hline
    \hline
    \multicolumn{1}{|l|}{\multirow{3}{*}{$f^\mathrm{cl}_\mathrm{PBH}=10^{-1}$}} & $M_\mathrm{PBH} = 1\ M_\odot$ & $5.0 \times 10^{-3}$ & $4.6 \times 10^{-4}$ & $1.1 \times 10^{-4}$ \\
    \multicolumn{1}{|l|}{} & $M_\mathrm{PBH} = 10\ M_\odot$ & $5.6 \times 10^{-3}$ & $3.5 \times 10^{-3}$ & $9.9 \times 10^{-4}$ \\
    \multicolumn{1}{|l|}{} & $M_\mathrm{PBH} = 100\ M_\odot$ & $9.9 \times 10^{-5}$ & $4.5 \times 10^{-4}$ & $1.2 \times 10^{-2}$ \\
    \hline
    \end{tabular}
    \caption{Local binary merger rate summed over direct capture and three-body processes for GCs with different PBH abundances and masses.}
\label{tab:merger_rate_densities}    
\end{table}

\noindent
We note that in most of the cases where we have PBHs with~$M_\mathrm{PBH}\lesssim 10\ M_\odot$, the total BH (i.e.,~summing over ABH-ABH, ABH-PBH, and PBH-PBH contributions) local merger rate is comparable with the ABH-ABH merger rate~$R^{\rm (DC+3B)}_{\rm ABH\ only} \simeq 5\times 10^{-3}\ \mathrm{Gpc^{-3}yr^{-1}}$ obtained considering GCs without PBHs.
While absolute numbers always depend on the model implemented and they should be taken with caution, this trend seems to suggest that there might exist scenarios where the knowledge of merger rates alone is not sufficient at all in discriminating between radically different alternatives.

\begin{figure}[ht]
    \centerline{
    \includegraphics[width=\columnwidth]{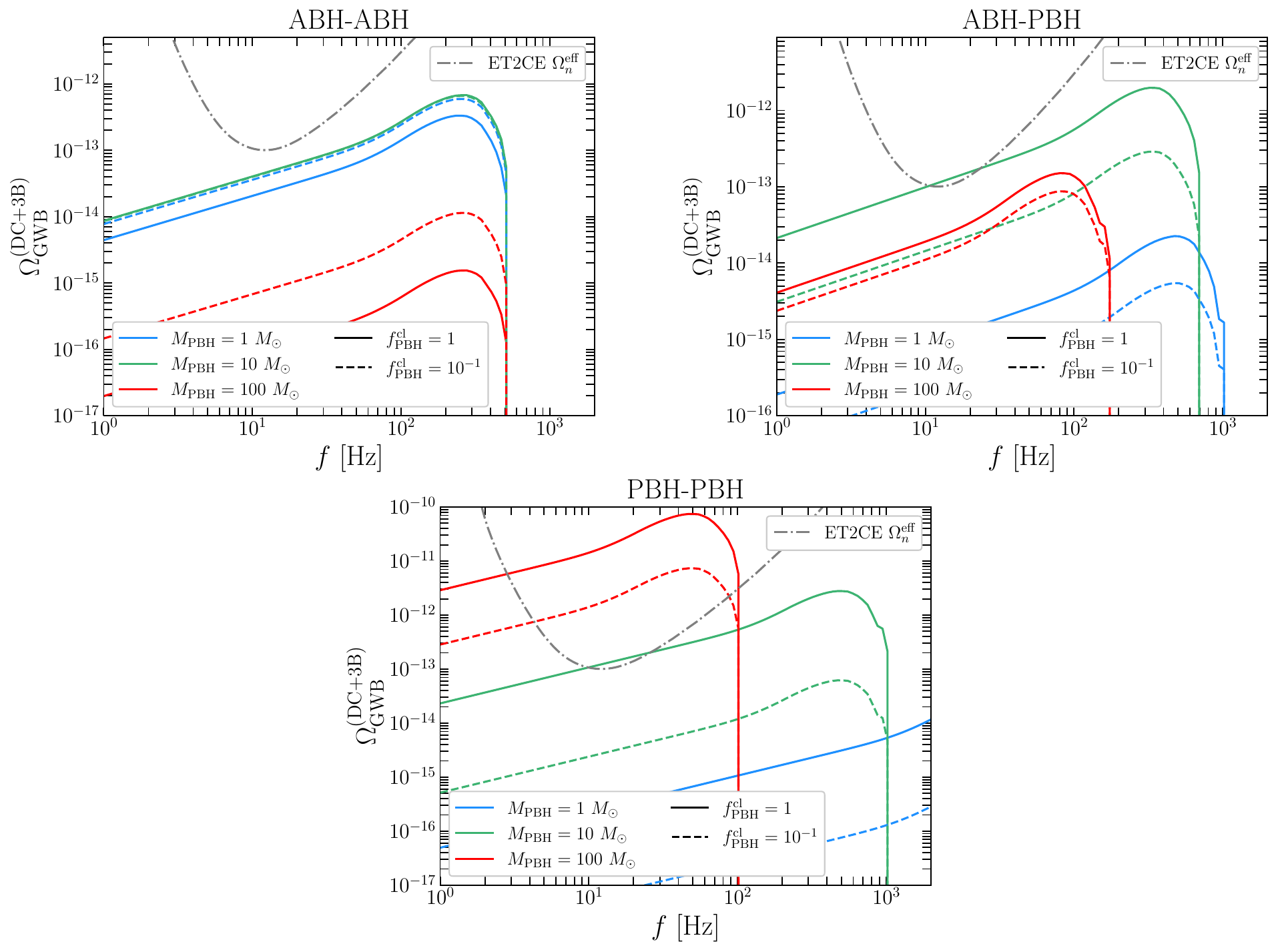}}
    \caption{GWB generated by ABH-ABH (\textit{upper left panel}), ABH-PBH (\textit{upper right panel}), and PBH-PBH (\textit{lower panel}) binaries in GCs.
    In each panel \textit{solid} and \textit{dashed lines} indicate~$f_{\rm PBH}^{\rm cl}=1$ and~$f_{\rm PBH}^{\rm cl}=10^{-1}$ PBH abundances, respectively.
    Darker shades of \textit{red}, \textit{green} and \textit{blue lines} indicate heavier PBH masses in GCs.
    The \textit{gray dotted-dashed line} represents the forecasted sensitivity of the ET2CE GW detector network.}
    \label{fig:GWB_binary_formation}
\end{figure}

These systems emits in the Hz-kHz frequency band, which is targeted by current and future ground-based interferometers.
On the observational side, there are already experimental upper limits on the GWB amplitude~\cite{abbott:GWBfirstrun, abbott:GWBsecondrun, abbott:GWBthirdrun} in this band, and a detection is within reach during next observational runs~\cite{regimbau:GWBdetectionforecast, abbott:GWBdetectionforecast}.
However, in the following figures we include only the effective ``sensitivity'' curves~\cite{thrane:gwdetectionstrategy, bartolo:lisaanisotropies} for a GW detector network given by Einstein Telescope and two Cosmic Explorers (ET2CE).
The sensitivity is determined by the expected noise level for the experiments, and it accounts for one year of observation time.
These effective sensitivity curves~$\Omega^\mathrm{eff}_n(f)$ are a useful benchmark for comparison: any GWB with a power-law-like frequency dependence (as in most of our cases) that intersects these curves is expected to be detected with a signal-to-noise ratio larger than unity after a year of observation. 
In figure~\ref{fig:GWB_binary_formation}, we show the GWB generated by ABH-ABH, ABH-PBH, and PBH-PBH systems in GCs with different PBH masses and abundances.
We do not expect ABH-ABH and ABH-PBH contributions to be detectable.
On the other hand, if PBHs have masses~$M_\mathrm{PBH} \gtrsim 10\ M_\odot$, we observe that detection prospects for the PBH-PBH component are in fact quite good, even for abundances as low as~$f^\mathrm{cl}_\mathrm{PBH} \simeq 10^{-1}$ in the most massive case considered here.
We stress again that the comparison between the benchmark effective sensitivity curves and our GWB estimates is provided only to visualize the relative importance of the different components, and to understand which ones can provide a detectable signal.
A complete study on the detectability of these contributions, as well as the possibility of separate them, is left for future work.

\subsection{Signatures of primordial black hole binaries}
\label{subsec:pbh_binary_signatures}

We showed in the previous section that in many circumstances PBH binaries formed in dense environments leave a detectable contribution in the GWB.
It is thus worth investigating whether this imprint is unique or not.
Moreover, PBHs form binaries via many different formation mechanisms, hence we necessarily have to consider all of them for our analysis to be comprehensive.
In particular, although each source contributing to the GWB evolves independently from the others, the joint set of channels has to be compatible with the local GW merger rate~$R^\mathrm{LVK}_0 \simeq 20 \, \mathrm{Gpc^{-3} \, yr^{-1}}$ measured by the LIGO-Virgo-KAGRA collaboration~\cite{abbott:gwtc3properties}, i.e., 
\begin{equation}
    R^\mathrm{LVK}_0 = R^\mathrm{EPBH}_0 + R^\mathrm{LPBH}_0 + R^\mathrm{ABH}_0 + R^\mathrm{GC}_0 \, ,
\label{eq:local_merger_rates}
\end{equation}
where the sum includes the early-time primordial, late-time primordial, astrophysical, and dense environments contributions, respectively.
The dense environment contribution contains PBH-PBH, ABH-PBH, and ABH-ABH binaries.

The rates for early- and late-time PBH binary formation are computed as discussed in sections~\ref{subsec:early_time_pbh_binaries} and~\ref{subsec:late_time_pbh_binaries}.
On the one hand, the late-time PBH binary merger rate is independent of PBH mass (for the mass ranges considered here), and it is given by~$R_0^{\rm LPBH}(f_\mathrm{PBH}=1) = 1.9 \, {\rm Gpc}^{-3} \, {\rm yr}^{-1}$~\cite{bosi:gwxlss,scelfo:gwxlssI,bird:pbhlatetimebinaries}.
As for early binaries, even though some initial theoretical estimates suggested large merger rates, recent numerical results suggest that such values overestimate the real one~\cite{Jedamzik:2020ypm, Jedamzik:2020omx}.
In a similar fashion to ref.~\cite{bosi:gwxlss}, the parameter $\mathcal{A}_m$ is defined for each fixed value of~$M_{\rm PBH}$ in such a way that, when~$f_{\rm PBH}=1$, the total contribution from PBH binaries in the field constitutes a certain fraction of the local GW merger rate~\cite{Franciolini:2021tla}.
As illustrative examples, we consider the cases in which~$R^\mathrm{EPBH}_0+R^\mathrm{LPBH}_0 = 0.1 \times R^\mathrm{LVK}_0$ and ~$R^\mathrm{EPBH}_0+R^\mathrm{LPBH}_0 = 0.3 \times R^\mathrm{LVK}_0$.\footnote{This value was suggested by some previous studies~\cite{Franciolini:2021tla}, and we consider it an upper limit for the fraction of LVK events that could be primordial.}
With these requirements, for~$M_{\rm PBH} = 100 \, M_{\odot}$, we set the early-time amplitude parameter to~$\mathcal{A}_m(f_\mathrm{PBH}=1) = 12 \, {\rm Gpc}^{-3} \, {\rm yr}^{-1}$ for a~$30\%$ contribution, or~$\mathcal{A}_m(f_\mathrm{PBH}=1) = 0.37 \, {\rm Gpc}^{-3} \, {\rm yr}^{-1}$ for a~$10\%$ contribution.

As we discussed in section~\ref{sec:PBH}, while the early- and late-time PBH binaries merger rate depends on the~$f_\mathrm{PBH}$ abundance parameter, the dense environment contribution depends on the cluster parameter~$f^\mathrm{cl}_\mathrm{PBH}$. As we showed in section~\ref{subsec:globular_cluster_number_density}, as long as~$f_\mathrm{PBH} \gtrsim 2\times 10^{-5}$ the value of the parameter $f_{\rm PBH}^{\rm cl}$ can be chosen independently from it. In other words, if PBHs constitute a small fraction of the dark matter, the contribution of the late and early formation channels to the GWB decreases; on the other hand, even if the contribution of the binaries in the cluster is small, it remains constant independently from the global $f_{\rm PBH}$, once that the abundance inside the cluster is fixed through $f_{\rm PBH}^{\rm cl}$. Therefore, there are regimes in which the relative contribution to the GWB of PBHs in dense environments is more relevant than the one of the early and late formation channels.

\begin{figure}[ht]
    \centerline{
    \includegraphics[width=.85\columnwidth]{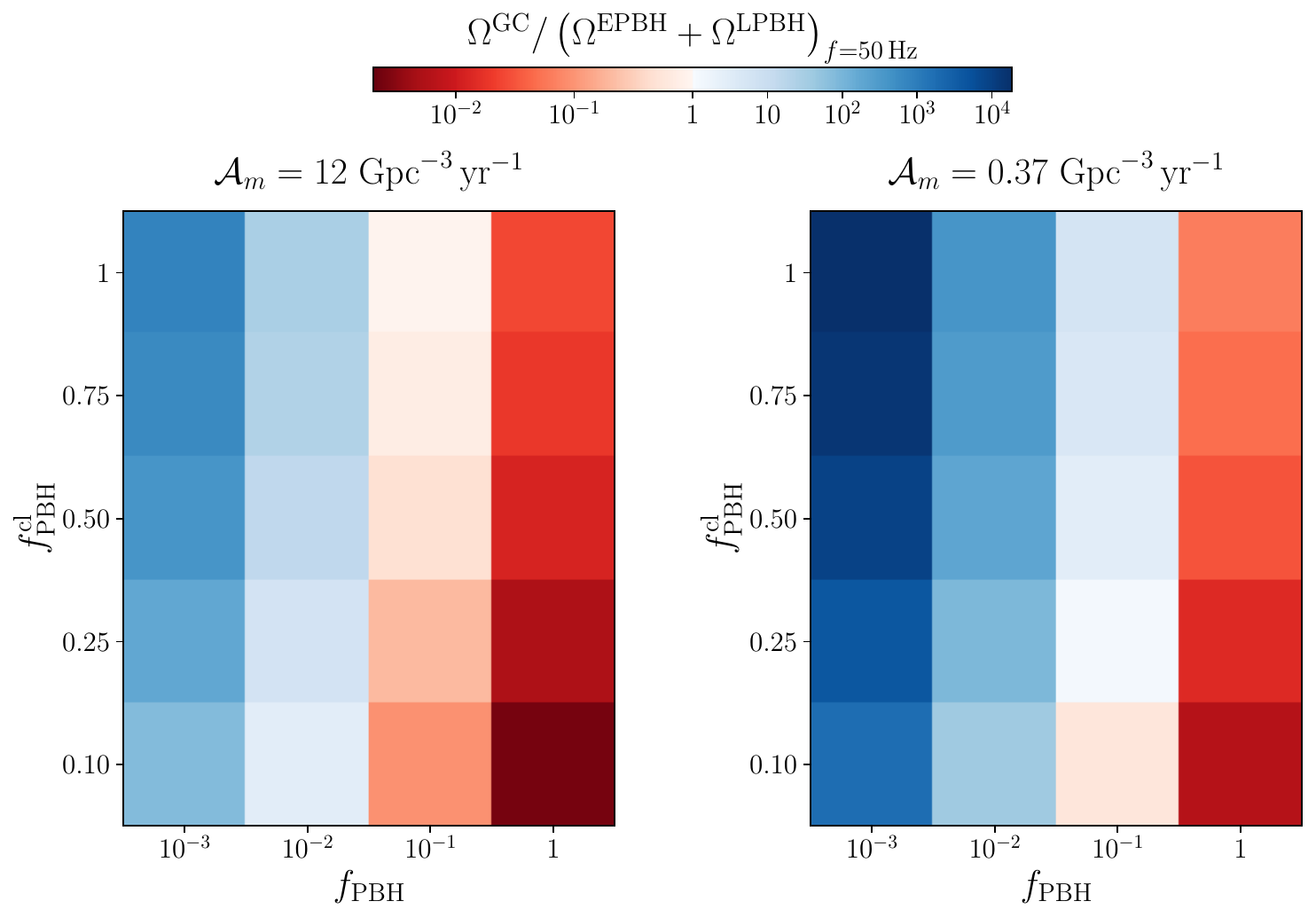}}
    \caption{Ratio of the GWB from GC to the total GWB expected from early- and late-time PBH binaries. The ratio is computed at the reference frequency~$f=50 \, {\rm Hz}$ where the PBH backgrund is expected to peak for~$M_{\rm PBH} = 100 \, M_{\odot}$.
    The two panels showcase different scenarios, where the total signal from early and late PBH binaries with~$f_{\rm PBH}=1$ contributes a~$30\%$ (\textit{left panel}) and~$10\%$ (\textit{right panel}) of the total merger rate.
    In the \textit{left panel},~$\mathcal{A}_m = 12 \, {\rm Gpc}^{-3} \, {\rm yr}^{-1}$, and the resulting early-time binaries local merger rate is~$R_0^{\rm EPBH}(f_{\rm PBH}=1) = 4 \, {\rm Gpc}^{-3} \, {\rm yr}^{-1}$. In the \textit{right panel},~$\mathcal{A}_m = 0.37 \, {\rm Gpc}^{-3} \, {\rm yr}^{-1}$, and~$R_0^{\rm EPBH}(f_{\rm PBH}=1) = 0.1 \, {\rm Gpc}^{-3} \, {\rm yr}^{-1}$. In both cases, the late-binaries local merger rate is~$R_0^{\rm LPBH}(f_{\rm PBH}=1) = 2 \, {\rm Gpc}^{-3} \, {\rm yr}^{-1}$.
    }
    \label{fig:2Dplot}
\end{figure}

Figure~\ref{fig:2Dplot} shows the relative importance of the GC contribution with respect to the total PBH contribution from the field, with varying~$f_{\rm PBH}$ and~$f_{\rm PBH}^{\rm cl}$.
The figure shows the ratio of the GWB from GC to the sum of the early- and late-time contributions, i.e.,~$\Omega^{\rm GC} / \left( \Omega^{\rm EPBH}+\Omega^{\rm LPBH} \right)$, computed at the reference frequency~$f=50 \, {\rm Hz}$ where the PBH background is expected to peak for~$M_{\rm PBH} = 100 \, M_{\odot}$.
Depending on the value of both~$f_{\rm PBH}$ and~$f_{\rm PBH}^{\rm cl}$, the GC contribution may be larger than the one that is usually estimated in the literature from early- and late-time PBH binaries. 

\begin{figure}[ht]
    \centerline{\includegraphics[width=\columnwidth]{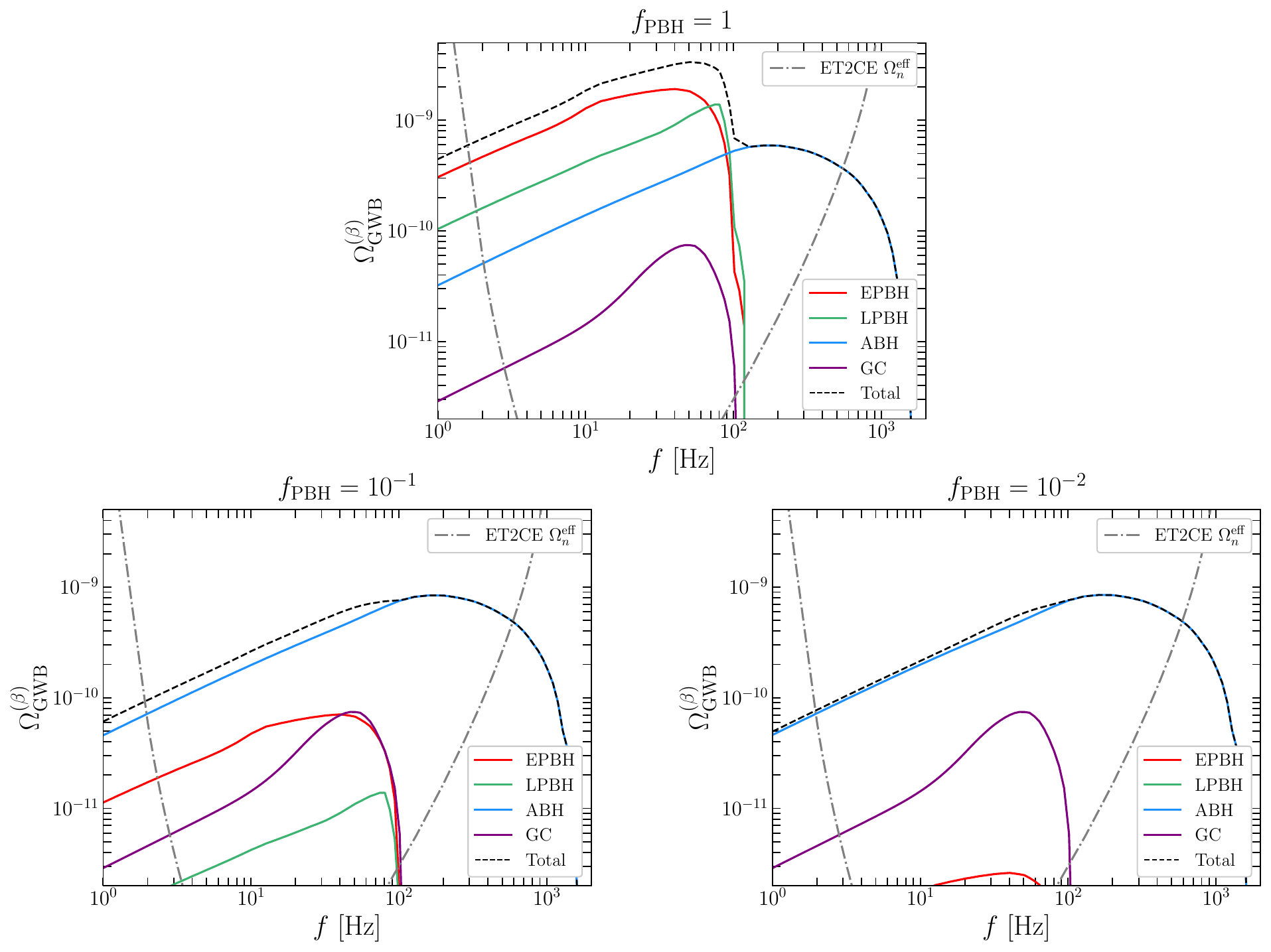}}
    \caption{GWB generated by early-time PBH (\textit{red lines}), late-time PBH (\textit{green lines}), ABH (\textit{blue lines}), and total GC (\textit{purple lines}) binaries when the PBH abundance is set to~$f_{\rm PBH} = 1$ (\textit{upper panel}), $f_{\rm PBH} = 10^{-1}$ (\textit{lower left panel}), and~$f_{\rm PBH} = 10^{-2}$ (\textit{lower right panel}).
    In all cases PBHs have~$M_\mathrm{PBH} = 100\ M_\odot$, and we set~$\mathcal{A}_m$ following the procedure outlined in the main text.
    The contribution from GCs assumes~$f^\mathrm{cl}_\mathrm{PBH}=1$.
    The sum of all contribution is represented by a \textit{black dashed line}.
    The ET2CE effective sensitivity curve for one year of observation time is showed as a \textit{dotted-dashed gray line}.}    \label{fig:pbh_signature}
\end{figure}

This relative importance must be considered in a wider scenario, that also accounts for the other contributions to the total GWB.
We show in figure~\ref{fig:pbh_signature} the GWB generated by early-time, late-time, ABH, and GC binaries when~$M_\mathrm{PBH}=100\ M_\odot$ and~$f_\mathrm{PBH} = 1, 10^{-1}, 10^{-2}$.
In the first case, the contribution from early- and late-time PBHs constitute a~$30\%$ of the total local binary merger rate -- the other cases are rescaled according to~\eqref{eq:PBHrate_early} and~\eqref{eq:PBHrate_late}. 
While in the~$f_{\rm PBH}=1$ case the sum of the early- and late- binary contributions outnumbers the contribution PBHs inside clusters, when~$f_{\rm PBH}=10^{-1}$ the contributions are comparable, and for~$f_{\rm PBH}=10^{-2}$, PBH binaries inside dense environment contribute to the GWB about one order of magnitude more than the other channels.
These scenarios showcase how even if~$f_\mathrm{PBH} \ll 1$, the GC contribution can leave a order tens-percent feature on the total GWB, especially in those cases where PBHs have masses larger than those of ABHs\footnote{This conclusion relies on a single effective mass to describe the ABH population. The implications of an extended mass function will be investigated in a future work.}, since the characteristic frequency of the merger scales as~$f \propto M^{-1}$. 
In other words, the presence of GWB features, which persist even in the case PBHs contribute only to a fraction of the dark matter, can potentially provide an indicator of more exotic binary formation channels, and deserves additional scrutiny.

As an illustrative example, in figure~\ref{fig:pbh_signature} we also show the ET2CE effective sensitive curve for one year of observation time, introduced in section~\ref{subsec:total_binary_contribution_from_gc}.
These detectors will likely detect the SGWB, but their observation will account for the overlapping signals from all the contributions (ABHs, early PBHs, late PBHs, PBHs from dense environments, other sources).
Disentangling them will require advanced component separation techniques (e.g.,~ref.~\cite{parida2019componentseparationmapmakingstochastic}); the detectability of the PBH contribution from clusters, as well as from early-time and late-time binaries, can be properly understood only through the development of a full analysis in this sense, which is beyond the scope of this paper, and is left to future work.

The GC contribution to the SGWB suffers from uncertainties related to the astrophysical modeling, which is still an object of debate.
Throughout our analysis, in order to keep the modeling simple, we have fixed the astrophysical parameters to values that are usually accepted in the literature.
We found that a major source of variability in the properties that determine the SGWB is the concentration parameter or (dimensionless) central gravitational potential, $\Psi_0/\sigma^2$, introduced in section~\ref{subsec:globular_cluster_modeling}, that roughly controls the central density of the cluster.
Typical values are reported to be~$\Psi_0/\sigma^2 \sim \mathcal{O}(10)$~\cite{gunn:globularclusters, oleary:hardnessratio,miocchi:energyequipartition}.
As an example, a population of more concentrated clusters, $\Psi_0/\sigma^2=12$, would be characterized by larger core densities, but lower segregation volumes. In this case, the two effects compensate in such a way that the total contribution from GCs to the SGWB is left unchanged. However, for a population of more diluted clusters, $\Psi_0/\sigma^2=8$, the two effects do not balance out, and the lower number densities result in a~$\sim 1/3$ decrease in the SGWB.
In order to assess the variability of the concentration parameter, and the impact of the most extreme cases on the final SGWB from GCs, a more complete characterization of the GC population will be needed.

Finally, we briefly comment on the choice of PBH mass function.
Throughout our analysis, we assumed that all PBHs have the same mass, namely we considered a monochromatic mass distribution in which~$dn_\mathrm{PBH}/dM_\mathrm{PBH} = \delta^D(M_\mathrm{PBH}-M^\star_\mathrm{PBH})$.
This choice remains the most commonly adopted in the literature, and it is always implicit in summary plots constraining~$f_{\rm PBH}$, compare e.g.,~with ref.~\cite{carr:pbhconstraintsreview}.
Even if~$f_{\rm PBH}=1$ is ruled out for most mass windows, smaller PBH density fractions are still allowed; for example, in the~$M_{\rm PBH}=\{1,10,100\}\,M_\odot$ cases discussed in the main text,~$f_{\rm PBH}>\{10^{-2},10^{-1},10^{-7}\}$ is ruled out at~$1\sigma$.
There are instead no constraints for smaller values of~$f_{\rm PBH}$, for which we showed that clusters can still contribute non-negligibly to the SGWB.

Increasing attention has been given to the extended mass function scenario, since it provides a reasonable and physically-motivated way out from the previously mentioned constraints: if the PBH mass can vary, $f_{\rm PBH}=1$ can still be possible. Various techniques have been suggested to convert the~$f_{\rm PBH}$ constraints obtained for monochromatic mass functions to extended scenarios, without re-analysing the data~\cite{Carr_2016,Carr_2017_ext,bellomo:pbhconstraints}.
In the context of this work, however, these techniques cannot be directly applied.
On the other hand, our formalism can be straightforwardly expanded to account for an extended PBH mass function~$dn_\mathrm{PBH}/dM_\mathrm{PBH}$.
To do so, it is sufficient to add~$M_{\rm PBH}$ to the parameter set~$\boldsymbol{\theta}$ and~$d\Phi_\mathrm{PBH}/dM_\mathrm{PBH}$ to the set of~$p(\boldsymbol{\theta},z)$ that enter equation~\eqref{eq:omega_sgwb}.
This leads to a consistent estimate of the SGWB, but it increases the computational cost of the analysis.
Moreover, this choice complicates the internal structure of the cluster, which can no longer be approximated by a three-population model.
A possible way out is to estimate an effective mass for the PBHs from their extended mass function, similarly to what has been done for the ABH case in equation~\eqref{eq:MeffABH}.

To check how this would affect our results, we consider a power-law mass function\footnote{The power-law mass function well describes scenarios in which PBHs are formed from the collapse of large density fluctuations~\cite{CARR1957} or cosmic strings~\cite{HAWKING1989237}.}
\begin{equation}
    \frac{dn_\mathrm{PBH}}{dM_\mathrm{PBH}} = \frac{\mathcal{N_{\rm PL}}}{M_{\rm PBH}^{1-\gamma}}\Theta_H(M_{\rm PBH}-M_{\rm min})\Theta_H(M_{\rm max}-M_{\rm PBH}) \, , 
\end{equation}
where~$\Theta_H(M_1-M_2)$ is the Heaviside Theta function, and the normalization factor is found solving
\begin{equation}
\begin{gathered}
    \rho_{c}^{\rm PBH} = \int dM_{\rm PBH} \frac{dn_{\rm PBH}}{dM_{\rm PBH}} M_{\rm PBH}  = \int_{M_{\rm min}}^{M_{\rm max}} dM_{\rm PBH} \mathcal{N}_{\rm PL} M_{\rm PBH}^{\gamma} \, , \\
    \mathcal{N}_{\rm PL} = \rho_{c}^{\rm PBH} \left[ \int_{M_{\rm min}}^{M_{\rm max}} dM_{\rm PBH} \, M_{\rm PBH}^{\gamma} \right]^{-1} \, .
\end{gathered}
\end{equation}
Ref.~\cite{bellomo:pbhconstraints}, showed that typical parameters for the power law are~$\gamma = 0,\,M_{\rm min} = 1\,M_\odot,\,M_{\rm max}=200\,M_\odot$ (which includes the cases~$M_{\rm PBH} = \{1,10,100\}\,M_\odot$ we analysed in the main text).
Then, the analogous of equation~\eqref{eq:MeffABH} for the PBH effective mass provides~$M_{\rm PBH}^{\rm eff} \simeq 102\,M_\odot$ for PBH-PBH mergers sourced by direct capture events.
The effective mass for ABH-ABH processes, as well as three-body and hyperbolic encounters, can be obtained analogously.
This effective mass is very close to the monochromatic~$M_{\rm PBH}=100\,M_\odot$ scenario depicted in figure~\ref{fig:pbh_signature}, hence we expect the results for the power-law mass function to be very close to the ones we previously discussed.
A similar calculation can be performed using different power laws or different extended mass functions.


\subsection{Hyperbolic encounters}
\label{subsec:hyperbolic_encounters}

Binary formation is not the only class of dynamical processes that take place in dense environments; in fact, we are guaranteed to have also a second class of events in GCs, i.e.,~GW bursts.
When two unbound compact objects pass close to each other, they emit gravitational radiation due to the reciprocal deflection of their orbit, analogously to the bremsstrahlung emission in the electromagnetic case~\cite{kovacs:gravitationalbremsstrahlung}. 
In this section, we are interested in those cases where the energy dissipated through GWs in the encounter is not sufficient to create a bound pair, i.e.,~both initial and final eccentricities are larger than unity.
These events are called hyperbolic encounters~($\beta = {\rm HE}$ hereafter), and they have been investigated for both ABHs~\cite{Kocsis_2006, capozziello:hyperbolicencounters, Mukherjee:2020hnm, Dandapat:2023zzn} and PBHs~\cite{garciabellido:pbhhyperbolicencountersI, garciabellido:pbhhyperbolicencountersII, morras:pbhhyperbolicencounters}.
Since multiple authors have provided slightly different descriptions of the hyperbolic encounter dynamics and GW emission, in appendix~\ref{app:hyperbolic_encounter_dynamics} we summarize the equations implemented in our code to model these interactions.

The hyperbolic encounter cross section reads as~\cite{capozziello:hyperbolicencounters}
\begin{equation}
    \sigma^{\rm (HE)} = \pi r^2_\mathrm{ini} \sin^2\theta_\mathrm{ini} ,
\end{equation}
where~$r_\mathrm{ini}$ is the distance between BHs at the beginning of the process, and~$\cos\theta_\mathrm{ini} = \hat{\mathbf{r}}_\mathrm{ini} \cdot \hat{\mathbf{v}}_\mathrm{rel}$ is the angle between the initial separation and relative velocity vectors.
In this case the velocity-averaged cross section reads as
\begin{equation}
    \left\langle \sigma^{(\mathrm{HE})} v_{\rm rel} \right\rangle = \int dr_\mathrm{ini} d\theta_\mathrm{ini} dv_\mathrm{rel} p(r_\mathrm{ini},\theta_\mathrm{ini},v_\mathrm{rel}) \sigma^{\rm (HE)} v_\mathrm{rel} \, ,
\end{equation}
where we derive for the first time the probability distribution functions for the initial distance, angle and relative velocity probability distribution function.
These are presented in appendix~\ref{app:hyperbolic_encounter_pdfs}.

In this scenario, it is more convenient to consider a single population of BHs instead of two different ones, see the discussion in appendix~\ref{app:hyperbolic_encounter_pdfs}.
Therefore, instead of computing the effective mass for ABHs, we compute a global effective mass~$M^\mathrm{eff}_\mathrm{HE}$ for the ensemble of ABHs and PBHs by solving the equation
\begin{equation}
    \Omega^{(\mathrm{HE})}\left( \frac{dn}{dM} = \frac{\rho_{0\rm ABH} + \rho_{0\rm PBH}}{M_\mathrm{HE}^\mathrm{eff}} \delta^D(M - M_\mathrm{HE}^\mathrm{eff}) \right) = \Omega^{(\mathrm{HE})} \left( \frac{dn_\mathrm{tot}}{dM} \right) \, ,
\end{equation}
where
\begin{equation}
    \frac{dn_\mathrm{tot}}{dM} = \frac{dn_{\rm ABH}}{dM_{\rm ABH}} + \frac{dn_{\rm PBH}}{dM_{\rm PBH}} = \frac{(2-\alpha) M^{-\alpha} \rho_{0\rm ABH}}{(M^\mathrm{max}_\mathrm{ABH})^{2-\alpha} - (M^\mathrm{min}_\mathrm{ABH})^{2-\alpha}} + \frac{\rho_{0\rm PBH}}{M_{\rm PBH}} \delta^{\rm D}(M-M_{\rm PBH}) \, ,
\end{equation}
and we use the fact that the GW strain scales as~$|h^{\mathrm{HE}}_{ij}|^2 \propto \mu^2 M^4$, as showed in equation~\eqref{eq:energy_hyperbolic}, for hyperbolic encounters.

\begin{figure}[ht]
    \centerline{
    \includegraphics[width=.6\columnwidth]{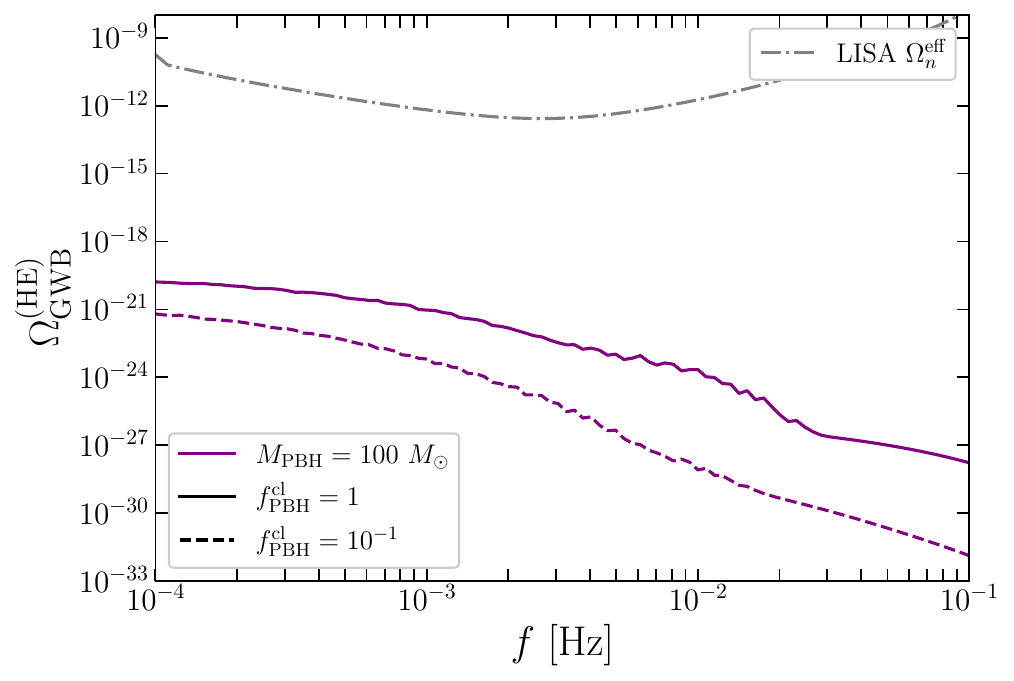}}
    \caption{Hyperbolic encounter GWB for PBH cluster abundances of~$f_{\rm PBH}^{\rm cl}=1$ (\textit{solid line}) and~$f_{\rm PBH}^{\rm cl}=10^{-1}$ (\textit{dashed line}). 
    In all cases reported we assume~$M_{\rm PBH}=100 \, M_{\odot}$. 
    The LISA effective sensitivity curve for one year of observation time is showed as a \textit{dotted-dashed gray line}.}
    \label{fig:GWB_HE}
\end{figure}

In this case, there is no time delay between the dynamical encounter and GW emission, hence in the computation of the GWB we only need to implement the interaction rate.
This class of dynamical interactions is extremely common, with interaction rates per cluster of order~$\Gamma^{(\rm HE)}_\mathrm{tot}(M_{\rm b}^{\rm ini} = 10^5 \, M_{\odot}) \simeq 10^{-5} \, {\rm yr}^{-1}$, approximately~$10^6$ times larger than the direct capture interaction rate per cluster.
However, as showed in figure~\ref{fig:GWB_HE}, the GWB produced by hyperbolic encounters turns out to be many orders of magnitude below LISA expected sensitivity.
The GWB generated by lighter PBHs has a similar magnitude to the ones reported in the figure.
Strong GW bursts are generated only by encounters with initial eccentricity very close to unity.
Unfortunately, that occurrence is very unlikely, as showed by the initial eccentricity probability distribution functions in appendix~\ref{app:hyperbolic_encounter_pdfs}, explaining the extremely weak GWB global signal.
Since the effective sensitivity scales with the square root of the observation time, we cannot foresee a detection of this class of events even for very massive PBHs.


\section{Conclusions}
\label{sec:conclusions}

In this work, we characterized the PBH contribution to the GWB, showing how dense environments provide the perfect ground for dynamical interaction that can ultimately be detected by current and next-generation GW observatories.
In particular, we showed that PBHs can leave a very characteristic imprint on the GWB even if their abundance is well below the current upper bounds.
Even in the case where the well-known GWBs from early- and late-time PBH binaries are below the sensitivity threshold of planned interferometers, PBHs in dense environments can still provide a contribution to the background that is comparable to the field-level ABH one, and therefore should not be neglected.
To understand how much the contribution of PBHs in clusters, as well as PBH early and late binaries, can affect the parameter inference from future data, a full component separation analysis is needed.
This, as well as a more accurate analysis of the detectability and a more in depth characterization of the variablity due to astrophysical properties, are left for a future work currently in preparation.

In this work we modeled the internal structure of GCs, following what appears to be the general consensus in the literature.
However, we are well aware that many properties of GCs are still an object of debate in the literature, and a change in any of them could significantly boost/reduce the dynamical interaction rate, and thus the GW signal.
As an example, a more accurate modeling of the signal, relying on numerical simulations, should also include GC evaporation and the subsequent evolution of the mass function~\cite{madrid2017:evaporation,Takahashi2001:dynamicalMF,Shin2008:dynamicalMF}, that we neglected at this stage, as well as an extended mass function for the ABHs.
Nevertheless, we believe that semi-analytical models, as the ones we propose, are still a valuable tool to explore the impact of astrophysical parameters and assess which regions should be further investigated numerically.

We showed that, if PBHs are present in the core of GCs, they have a significant impact on their dynamical properties and internal structure, causing the other populations to segregate within more or less diluted cores, depending on whether they are lighter or heavier than PBHs.
Therefore, properly characterizing the GWB contribution from PBHs in dense environments can also deepen our understanding of the internal structure of these astrophysical systems, and on the amount of dark matter within their cores. 

Moreover, even if in this paper we restricted our attention to GCs, we acknowledge that our formalism can be readily applied to other kinds of gravitationally bound system, such as nuclear clusters or even clusters made exclusively of PBHs~\cite{Belotsky:2018wph}.

In an upcoming work in preparation, we will study the possibility of disentangling the different channels and cases, so to investigate their detectability and the cosmological tests that this observable will enable.


\acknowledgments
The authors thank Konstantinos Kritos, Chiara Mingarelli, Emanuele Berti, Lucia Contiero for useful discussion. 

\noindent
SL thanks the John Hopkins University for the hospitality during the final stages of this work. The visit was supported by a  Balzan awarded scholarship, funded by the Balzan Centre for Cosmological Studies. SL is supported by an Azrieli International Postdoctoral Fellowship.
NB is supported by PRD/ARPE 2022 ``Cosmology with Gravitational waves and Large Scale Structure - CosmoGraLSS'' and acknowledges partial support from the National Science Foundation (NSF) under Grant No.~PHY-2112884.
AR acknowledges funding from the Italian Ministry of University and Research (MIUR) through the ``Dipartimenti di eccellenza'' project ``Science of the Universe''.
L.~V. acknowledges financial support from the Supporting TAlent in Re-Search@University of Padova (STARS@UNIPD) for the project “Constraining Cosmology and Astrophysics with Gravitational Waves, Cosmic Microwave Background and Large-Scale Structure cross-correlations”. 


\appendix
\section{Timescales}
\label{app:time_scales}

In this appendix we investigate different timescales of the systems of interest, i.e.,~GCs and binaries, and we discuss their relevance for the purpose of estimating the GWB generated by compact objects in dense environments.


\paragraph{Stellar evolution timescale.}
Even if at the time of formation we assume the GC to have exclusively a stellar population, massive stars are expected to rapidly evolve and form a population of ABHs.
In other words, the massive star and ABH mass functions are both evolving in time, with the former depleted in favor of the latter.
We estimate the ABH formation time as the minimum time required for a massive star to burn Hydrogen, Helium, Carbon and all the other elements in its core, i.e.,~$t_\mathrm{ABH} = t_\mathrm{H}+t_\mathrm{He,C,...}$, where the Hydrogen burning time is given by~\cite{kippenhahn:globularclustersimulations}
\begin{equation}
    t_\mathrm{H} \simeq 35.3 \left(\frac{8 \, M_\odot}{M_\star}\right)^{\alpha_\mathrm{H}-1} \ \mathrm{Myr} \, ,
\label{eq:time_abhformation}
\end{equation}
and the scaling exponent is~$\alpha_\mathrm{H}=3.8$. 
The burning time of heavier elements is of order~$t_\mathrm{He,C,...}\simeq 0.2\times t_\mathrm{H}$~\cite{kippenhahn:globularclustersimulations}. 
Therefore, we find that the minimum ABH mass inside in a GC is
\begin{equation}
    M^\mathrm{min}_\mathrm{ABH} = 
    \left\lbrace \begin{aligned}
        & 40\ M_\odot & \quad t \leq t_\mathrm{ABH}(40\ M_\odot) \, , \\
        & 8 \left[\frac{t_\mathrm{ABH}(8\ M_\odot)}{t}\right]^{\frac{1}{\alpha_\mathrm{H}-1}}\ M_\odot & \quad t_\mathrm{ABH}(40\ M_\odot) < t < t_\mathrm{ABH}(8\ M_\odot) \, , \\
        & 8\ M_\odot & \quad t \geq t_\mathrm{ABH}(8\ M_\odot) \, . \\
    \end{aligned} \right.
\label{eq_M_thr}
\end{equation}
Assuming that the GC and all stars inside it form at the same time, we have that after a typical maximum timescale of order~$t_\mathrm{ABH} \simeq 42 \, \mathrm{Myr}$ the ABH population is fully formed.
Given that this timescale is only a fraction of the typical segregation timescale (e.g.,~$t_\mathrm{ABH}/t_\mathrm{rel} \simeq 10^{-1}$, see following paragraphs), we can safely neglect the modelling of this transition time in our estimates.


\paragraph{Time delay between binary formation and merger.}
Mergers of BH binaries typically do not happen immediately after binary formation.
Instead, binaries present a typical time delay distribution determined by the initial properties of the system and of the environment.
Regarding direct capture processes, we follow ref.~\cite{cholis:orbitaleccentricities} and describe the initial state of a binary using its initial semi-major axis and eccentricity
\begin{equation}
    a_\mathrm{ini} = -\frac{G M\mu}{2E_{\rm ini}}, \qquad e_\mathrm{ini} = \sqrt{1+2\frac{E_{\rm ini} b^2 v_{\rm rel}^2}{G^2 M^2\mu}} \, ,
\end{equation}
respectively, where~$b$ is the impact parameter, $E_\mathrm{ini}$ is the energy of the binary at the beginning of the inspiral phase and it reads as
\begin{equation}
    E_{\rm ini} = \frac{\mu v_{\rm rel}^2}{2}-\frac{85\pi G^{7/2}}{12\sqrt{2}c^5}\frac{\mu^2 M^{5/2}}{r_p^{7/2}} \, ,
\end{equation}
and~$r_p$ being the distance of closest approach, given by
\begin{equation}
    r_p \simeq \frac{b^2 v_{\rm rel}^2}{2MG}\left(1-\frac{b^2v_{\rm rel}^4}{4M^2G^2}\right) \, .
\end{equation}
Therefore, the delay time between binary formation and merger reads as~\cite{peters:mergingtime}
\begin{equation}
    t_{\rm d} = \frac{3}{85}\frac{a_\mathrm{ini}^4c^5}{G^3 M^2 \mu} \left(1-e_\mathrm{ini}^2\right)^{7/2} \, .
\end{equation}
We create a mock catalogue of binaries formed through direct capture process and we derive their time delay distribution assuming that the impact parameter is uniformly distributed between its minimum and maximum values
\begin{equation}
    \begin{aligned}
        b_\mathrm{min} &= \sqrt{12} \frac{GM}{c v_\mathrm{rel}} \, , \qquad b_\mathrm{max} = \left(\frac{340\pi}{3}\right)^{1/7} \frac{GM^{6/7}\mu^{1/7}}{c^2} \left(\frac{v_\mathrm{rel}}{c}\right)^{-9/7} \, ,
    \end{aligned}
\end{equation}
respectively. 

\begin{figure}[ht]
    \centerline{
    \includegraphics[width=\columnwidth]{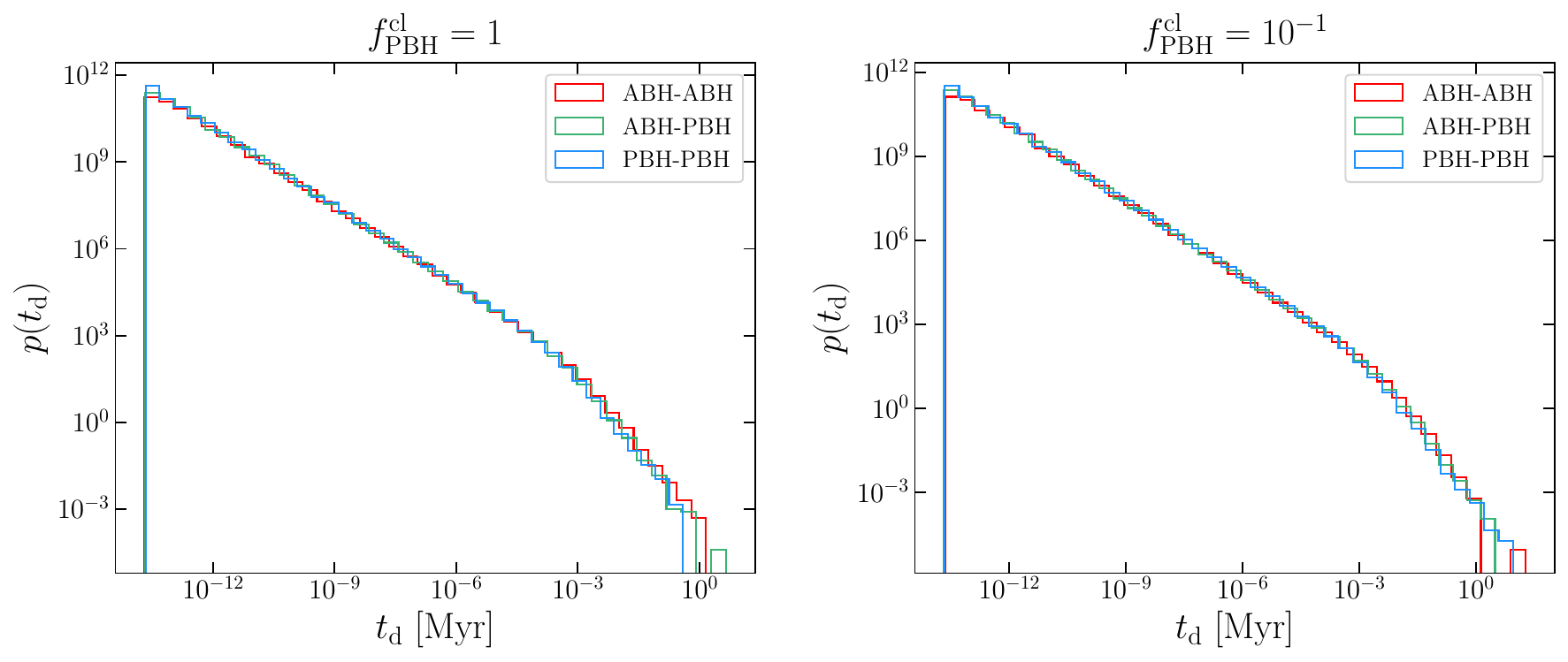}}
    \caption{Time delay probability distribution for ABH-ABH (\textit{red}), ABH-PBH (\textit{green}) and PBH-PBH (\textit{blue}) binaries for a GC with~$M^\mathrm{ini}_\mathrm{b} = 10^5 \, M_\odot$, $M_\mathrm{PBH} = 10 \, M_\odot$, and~$f^\mathrm{cl}_\mathrm{PBH}=1$ (\textit{left panel}) or~$f^\mathrm{cl}_\mathrm{PBH}=10^{-1}$ (\textit{right panel}).
    In the case of~$f^\mathrm{cl}_\mathrm{PBH}=1$ all the binaries have time delay at most of~$1 \, \mathrm{Myr}$, whereas in the~$f^\mathrm{cl}_\mathrm{PBH}=10^{-1}$ case a negligible fraction of events has also larger time delays ($0.03\%$, $0.01\%$ and~$0.05\%$ for ABH-ABH, ABH-PBH and PBH-PBH, respectively). }
    \label{fig:time_delay_distribution}
\end{figure}

We show the recovered time delay distributions in figure~\ref{fig:time_delay_distribution}, and we observe that for all classes of compact objects \textit{(i)} the time delay distribution scales approximately as~$p(t_{\rm d}) \propto t_{\rm d}^{-1}$, and \textit{(ii)} the maximum time delay is~$\mathcal{O}(1) \, \mathrm{Myr}$; therefore, since the average time delay is~$\left\langle t_{\rm d} \right\rangle \ll 1 \, \mathrm{Myr}$, we conclude that we can safely ignore this time delay in our modelling of the merger rate.
On the other hand, regarding the three-body encounters binary formation channel, we are not aware of any estimate on their expected time delay distribution.
However, we do not foresee the three-body time delay distribution to be significantly different from the direct capture one, since compact object participating in both kinds of interaction effectively populate the same environment.
Thus, also in the case of three-body encounters, we neglect the time delay between binary formation and merger.


\paragraph{Mass segregation timescale.}
The initial state of GCs are currently very uncertain, in particular little is known about whether they form already mass-segregated or not, see, e.g.,~refs.~\cite{Allison:2009ma, PortegiesZwart:2010cly} for a discussion on young globular clusters.
If they do~\cite{Klessen:2001ft, Bonnell:2006hk}, density and velocity profiles are expected to be well described by the phase space distribution in equation~\eqref{eq:king_phase_space_distribution} during almost all the GC lifetime.
In the opposite case, the picture presented in section~\ref{subsec:globular_cluster_modeling} is valid only after the GC relaxed into its ``equilibrium'' configuration, hence after a typical timescale from formation of the order of the relaxation time~$t_\mathrm{rel}$ defined in equation~\eqref{eq:trelax}.
In this work, we consider GCs that are not mass segregated at formation time.
However, since the typical relaxation time for GCs in the mass range of our interest is of order~$\mathcal{O}(300)\, \mathrm{Myr}$, and it is significantly lower than the typical timescale over which binaries form (see following paragraph), we neglect this effect in the modelling of the GW emission.
Note that even if very massive GC have a relaxation timescale of Gyrs, the steepness of the GC mass function render their contribution to the GWB subdominant.
Therefore, we do not expect this approximation to significantly alter our estimations.


\paragraph{Merger timescale.}
The mass evolution of the GC component might be affected by the very same merger events we are interested into.
If mergers happen frequently, i.e.,~if
\begin{equation}
    H/\Gamma^{(\beta)}_{ij,\rm tot} \ll 1 \, ,
\label{eq:interactionrate_over_hubbletime}
\end{equation}
the low-mass tail of the ABH and/or PBH mass function will get depleted of compact objects in favor of the high-mass tail.
Even if our initial assumption was that the mass function is constant in time, we check a posteriori that~$H_0/\Gamma^{(\beta)}_{ij,\rm tot} \gtrsim 1$ in all binary formation channels and for all combinations of sources.
Only in the case of GCs with~$M_\mathrm{PBH}=100 \, M_\odot$ and~$f^\mathrm{cl}_\mathrm{PBH}=1$ we observe ~$H_0/\Gamma^{(\mathrm{DC})}_{\rm PBH-PBH, tot} \simeq \mathcal{O}(\mathrm{few})$ when~$M^\mathrm{ini}_\mathrm{b} \gtrsim 10^6 \, M_\odot$, i.e.,~very massive GCs might experience few PBH-PBH binary mergers per Hubble time.
Since the GC mass distribution is dominated by GCs with~$M^\mathrm{ini}_\mathrm{b} \simeq 10^5 \, M_\odot$, we can safely conclude that mergers do not significantly alter the compact object mass function.


\section{Relative velocity probability distribution function}
\label{app:relative_velocity_pdf}

Cross sections of dynamical interactions described in section~\ref{sec:GWB} depend on the relative velocity of the compact objects involved in the process.
In this appendix we derive an approximate analytical expression for the relative velocity probability distribution function starting from the velocity distribution of BHs in the GC cores.

\begin{figure}[ht]
    \centerline{
    \includegraphics[width=\columnwidth]{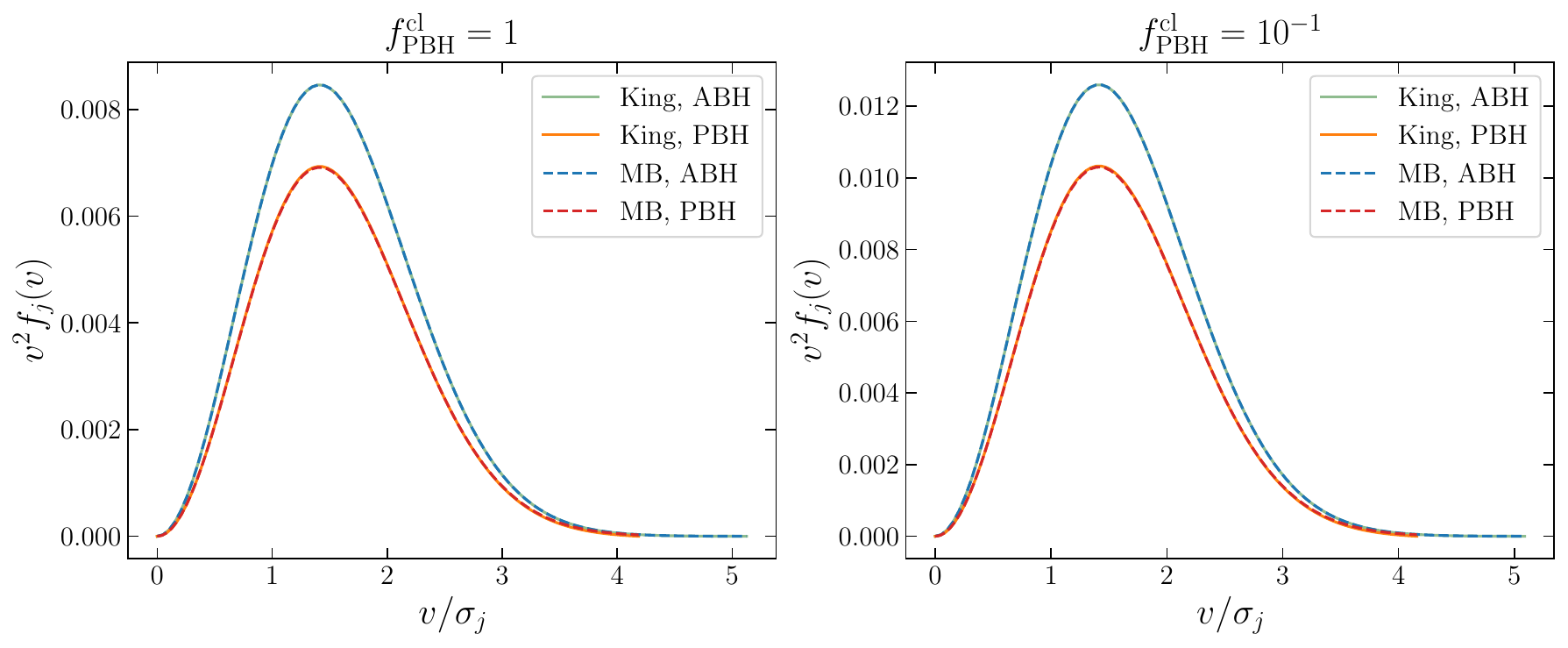}}
    \caption{King and truncated Maxwell-Boltzmann (MB) ABH and PBH velocity probability distribution functions in the core of GCs with~$M^\mathrm{ini}_\mathrm{b} = 10^5 \, M_\odot$, $M_\mathrm{PBH} = 10 \, M_\odot$, and~$f^\mathrm{cl}_\mathrm{PBH}=1$ (\textit{left panel}) or~$f^\mathrm{cl}_\mathrm{PBH}=10^{-1}$ (\textit{right panel}).
    Velocities on the x-axis are rescaled by each respective population velocity scale. }
    \label{fig:KMB_comparison}
\end{figure}

We compare in figure~\ref{fig:KMB_comparison} the velocity probability distribution function of individual BH populations given by equation~\eqref{eq:king_phase_space_distribution} to a truncated Maxwell-Boltzmann distribution
\begin{equation}
    p(\boldsymbol{v}_i) = \frac{\mathcal{N}_i}{(2\pi\sigma^2_i)^{3/2}} e^{-v_i^2/2\sigma^2_i} \, ,
\label{eq:p_b_MB}
\end{equation}
where the normalization factor reads as
\begin{equation}
    \mathcal{N}^{-1}_i = \frac{1}{(2\pi)^{3/2} \sigma^3_i} \int d\Omega_i \int_0^{v_\mathrm{esc}} dv_i\ v^2_i e^{-v_i^2/2\sigma^2_i} = \mathrm{erf}\left(\frac{v_\mathrm{esc}}{\sqrt{2} \sigma_i}\right) - \sqrt{\frac{2}{\pi}} \frac{v_\mathrm{esc}}{\sigma_i} e^{-v^2_\mathrm{esc}/2 \sigma^2_i} \, .
\end{equation}
As expected, since the typical velocity dispersion is smaller than the escape velocity, i.e.,~$\sigma_j \ll v_\mathrm{esc} = \sqrt{2\Psi_0}$ for each population of interest, we find that the truncated Maxwell-Boltzmann nicely approximates the real phase space distributions.
Since the difference between phase space distributions appears to be relatively small, in the following we adopt the truncated Maxwell-Boltzmann for our analytical estimates.

Under the change of variables defined by
\begin{equation}
    \boldsymbol{v}_\mathrm{rel} = \boldsymbol{v}_2 - \boldsymbol{v}_1 \, , \qquad \boldsymbol{v}_c = \frac{\sigma^2_2 \boldsymbol{v}_1 + \sigma^2_1 \boldsymbol{v}_2}{\sigma^2_\mathrm{rel}} \, ,
\end{equation}
we have that the original velocities read as 
\begin{equation}
    \boldsymbol{v}_1 = \boldsymbol{v}_c - \frac{\sigma^2_1}{\sigma^2_\mathrm{rel}} \boldsymbol{v}_\mathrm{rel} \, , \qquad \boldsymbol{v}_2 = \boldsymbol{v}_c + \frac{\sigma^2_2}{\sigma^2_\mathrm{rel}} \boldsymbol{v}_\mathrm{rel} \, , \\
\end{equation}
with~$\sigma^2_\mathrm{rel} = \sigma^2_1 + \sigma^2_2$ and~$\sigma^2_c = \sigma^2_1 \sigma^2_2 / (\sigma^2_1 + \sigma^2_2)$. 
Since the velocities of the two objects are bound to be lower than the escape velocity~$v_{\rm esc}$, the domain of the relative velocity is~$v_\mathrm{rel} \in [0, 2v_\mathrm{esc}]$ while~$v_{\rm c}$ has a ``circular'' domain, i.e.,
\begin{equation}
    v^2_c \leq v^2_{c,\mathrm{max}}(v_\mathrm{rel}) = \frac{\sigma^2_c}{\sigma^2_\mathrm{rel}} \left[ v^2_\mathrm{esc} \left(2 + \frac{\sigma^2_2}{\sigma^2_1} + \frac{\sigma^2_1}{\sigma^2_2} \right) - v^2_\mathrm{rel} \right] \, .
\end{equation}
The integration domain of~$\cos\theta_c = v_{c,z}/v_c$ is also restricted.
In particular, starting from the definitions of~$\cos\theta_c$ and $\cos\theta_\mathrm{rel} = v_{\mathrm{rel},z}/v_\mathrm{rel}$, we have
\begin{equation}
    \sigma^2_\mathrm{rel} v_c \cos\theta_c = \sigma^2_1 v_2 \cos\theta_2 + \sigma^2_2 v_1 \cos\theta_1 \, , \qquad \sigma^2_2 v_\mathrm{rel} \cos\theta_\mathrm{rel} = \sigma^2_2 v_2 \cos\theta_2 - \sigma^2_2 v_1 \cos\theta_1 \, .
\end{equation}
By summing these two equations and later squaring them, we find
\begin{equation}
    \cos^2\theta_c + 2 \frac{\sigma^2_2}{\sigma^2_\mathrm{rel}} \frac{v_\mathrm{rel}}{v_c} \cos\theta_\mathrm{rel} \cos\theta_c + \frac{\sigma^4_2}{\sigma^4_\mathrm{rel}} \frac{v^2_\mathrm{rel}}{v^2_c} \cos^2\theta_\mathrm{rel} = \frac{v^2_2}{v^2_c} \cos^2\theta_2 \leq \frac{v^2_\mathrm{esc}}{v^2_c} \, ,
\end{equation}
which admits solutions for~$\cos\theta_c$ in the range
\begin{equation}
    - \frac{\sigma^2_2}{\sigma^2_\mathrm{rel}} \frac{v_\mathrm{rel}}{v_c} \cos\theta_\mathrm{rel} - \frac{v_\mathrm{esc}}{v_c} \leq \cos\theta_c \leq - \frac{\sigma^2_2}{\sigma^2_\mathrm{rel}} \frac{v_\mathrm{rel}}{v_c} \cos\theta_\mathrm{rel} + \frac{v_\mathrm{esc}}{v_c} \, .
\end{equation}
When~$v_\mathrm{esc} \gg \sigma_1,\sigma_2$, as in our cases of interest, we have~$-1\lesssim \cos\theta_c \lesssim 1$, and the integration domain becomes approximately unconstrained.

\begin{figure}[ht]
    \centerline{
    \includegraphics[width=\columnwidth]{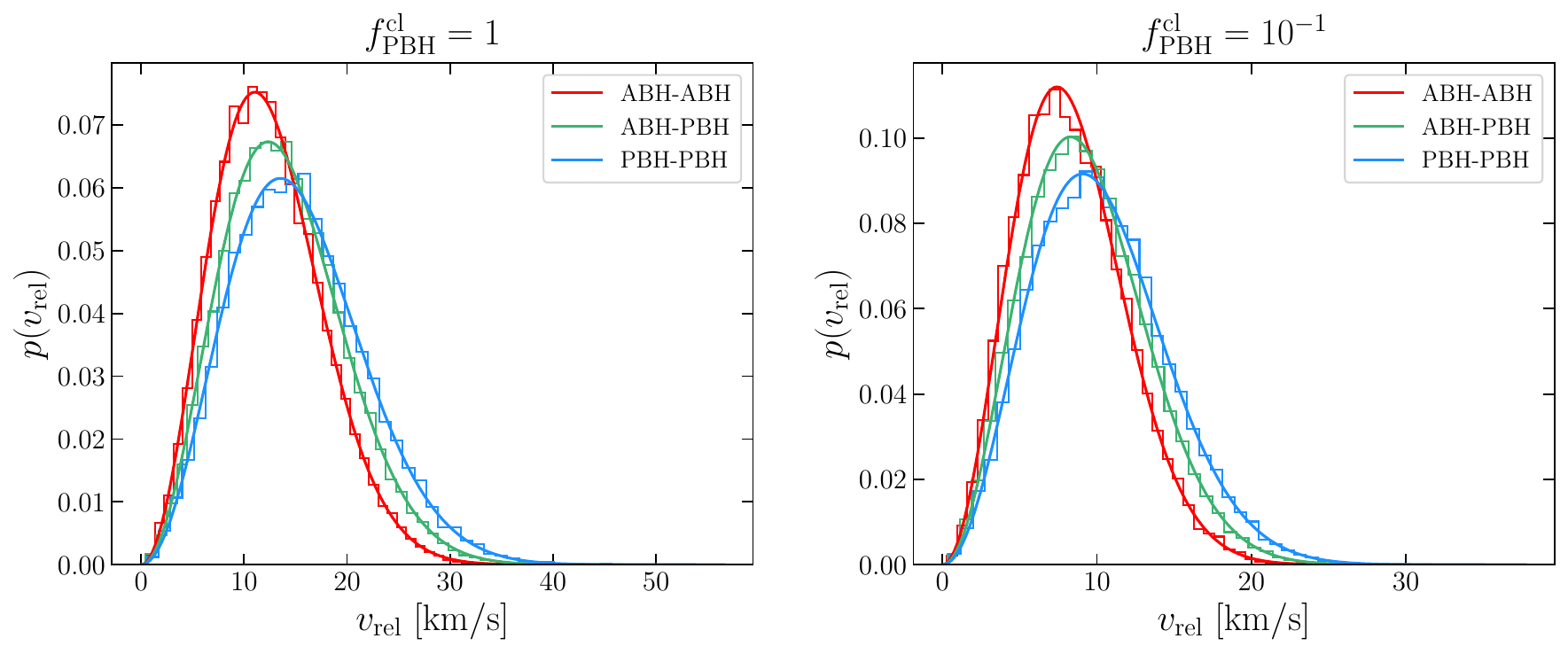}}
    \caption{Relative velocity probability distribution function obtain from a catalog (\textit{histograms}) compared to the analytic prediction of equation~\eqref{eq:relative_velocity_analytic_pdf} (\textit{solid lines}) for ABH-ABH (\textit{red}), ABH-PBH (\textit{green}) and PBH-PBH (\textit{blue}) pairs in a GC with~$M^\mathrm{ini}_\mathrm{b} = 10^5 \, M_\odot$, $M_\mathrm{PBH} = 10 \, M_\odot$, and~$f^\mathrm{cl}_\mathrm{PBH}=1$ (\textit{left panel}) or~$f^\mathrm{cl}_\mathrm{PBH}=10^{-1}$ (\textit{right panel}). }
    \label{fig:pdf_vrel_comparison}
\end{figure}

The probability distribution function transform under such change of variables as
\begin{equation}
    p(\boldsymbol{v}_1)p(\boldsymbol{v}_2) d^3v_1 d^3v_2 = p(\boldsymbol{v}_c,\boldsymbol{v}_\mathrm{rel}) d^3v_c d^3v_\mathrm{rel} = \frac{\mathcal{N}_1 \mathcal{N}_2}{(2\pi)^3 \sigma^3_c \sigma^3_\mathrm{rel}} e^{-v^2_c/2\sigma^2_c} e^{-v^2_\mathrm{rel}/2\sigma^2_\mathrm{rel}} d^3v_c d^3v_\mathrm{rel} \, ,
\end{equation}
therefore, when~$v_\mathrm{esc} \gg \sigma_1,\sigma_2$, we find that the relative velocity probability distribution function~$p(v_\mathrm{rel})$ reads as
\begin{equation}
    \begin{aligned}
        p(v_\mathrm{rel}) &= \mathcal{N}_1 \mathcal{N}_2 v^2_\mathrm{rel} \int d\Omega_\mathrm{rel} \int_0^{v_{c,\mathrm{max}}(v_\mathrm{rel})} d\Omega_c dv_c v^2_c \frac{1}{(2\pi)^3 \sigma^3_c \sigma^3_\mathrm{rel}} e^{-v^2_c/2\sigma^2_c} e^{-v^2_\mathrm{rel}/2\sigma^2_\mathrm{rel}} \\
        &= \mathcal{N}_1 \mathcal{N}_2 \frac{(4\pi)^2 v^2_\mathrm{rel} e^{-v^2_\mathrm{rel}/2\sigma^2_\mathrm{rel}}}{(2\pi)^3 \sigma^3_c \sigma^3_\mathrm{rel}} \int_0^{v_{c,\mathrm{max}}(v_\mathrm{rel})} dv_c v^2_c  e^{-v^2_c/2\sigma^2_c} \\
        &= \mathcal{N}_1 \mathcal{N}_2 \frac{(4\pi)^2 v^2_\mathrm{rel} e^{-v^2_\mathrm{rel}/2\sigma^2_\mathrm{rel}}}{(2\pi)^3 \sigma^3_\mathrm{rel}} \left[ \sqrt{\frac{\pi}{2}} \mathrm{erf} \left(\frac{v_{c,\mathrm{max}}(v_\mathrm{rel})}{\sqrt{2}\sigma_c} \right) - \frac{v_{c,\mathrm{max}}(v_\mathrm{rel})}{\sigma_c} e^{-v^2_{c,\mathrm{max}}(v_\mathrm{rel})/2\sigma^2_c} \right] \, . \\
    \end{aligned}
\label{eq:relative_velocity_analytic_pdf}
\end{equation}

In figure~\ref{fig:pdf_vrel_comparison} we show the comparison between the relative velocity distribution given by the theoretical prediction of equation~\eqref{eq:relative_velocity_analytic_pdf}, and by generating a catalog of events in which the velocities of the two bodies are separately sampled from the individual Maxwell-Boltzmann distributions.
For both benchmark cases, we find that our analytic formula is accurate at percent level, with deviations appearing only in the high relative velocity tail of the probability distribution.


\section{Hyperbolic encounter dynamics}
\label{app:hyperbolic_encounter_dynamics}

In this appendix we briefly summarize the dynamics of the hyperbolic encounter between two bodies, mainly following refs.~\cite{devittori:hyperbolicencounters, grobner:hyperbolicencounters}. 
In the rest frame of the second body, the coordinates of the first one are parametrized as
\begin{equation}
    x(u) = a(e - \cosh u) \, , \qquad y(u) = a (e^2-1)^{1/2} \sinh u \, ,
\end{equation}
where~$a=GM/v^2_\mathrm{rel}$ is the semi-major axis, $e$ is the eccentricity and~$u$ is the anomaly.
The latter implicitly parametrizes also the time evolution of the system as
\begin{equation}
    t(u) = \sqrt{\frac{a^3}{GM}} (e\sinh u - u) \, ,
\end{equation}
with~$t_0 = GM/v^3_\mathrm{rel}$ being a characteristic time of the gravitational interaction.
At time~$t=0$ (corresponding to $u=0$) the two bodies are at the minimal distance~$r_\mathrm{min}=a(e-1)$.

The power emitted in GWs reads as
\begin{equation}
    P(t) = \frac{G}{45c^5} \partial^3_t D_{ij}\partial^3_t D_{ij} \, ,
\end{equation}
where the quadrupole moment tensor is given by~$D_{ij} = 3M_{ij} - \delta_{ij} M_{kk}$ and the second mass moment is defined as
\begin{equation}
    M_{ij} = \frac{1}{c^2} \int d^3x\ T^{00}(\boldsymbol{x}) x_i x_j \, .
\end{equation}
In the simple case at hand we have~$T^{00}(\boldsymbol{x}) = \mu c^2 \delta^{D}\left( \boldsymbol{x}-\boldsymbol{x}(t) \right)$, therefore the energy emitted though GWs during the encounter reads as
\begin{equation}
    \Delta E = \int_0^\infty df \frac{dE}{df} = \frac{128\pi^6}{45} \frac{G^5 \mu^2 M^4}{c^5 v^8_\mathrm{rel}} \int_0^\infty df f^4 \mathcal{F}(2\pi t_0 f) \, ,
\label{eq:energy_hyperbolic}
\end{equation}
where
\begin{equation}
    \begin{aligned}
        \mathcal{F}(x) &= \left| \frac{3(e^2-1)}{e} H_{ix}^{(1)'}(ix e) + \frac{e^2-3}{e^2} \frac{i}{x} H_{ix}^{(1)}(ix e) \right|^2 \\
        &\qquad + \left| \frac{3(e^2-1)}{e} H_{ix}^{(1)'}(ix e) + \frac{2e^2-3}{e^2} \frac{i}{x} H_{ix}^{(1)}(ix e) \right|^2 \\
        &\qquad\qquad + \frac{18(e^2-1)}{e^2} \left| \frac{1}{x} H_{ix}^{(1)'}(ix e) + \frac{e^2-1}{e} i H_{ix}^{(1)}(ix e) \right|^2 \, ,
    \end{aligned}
\end{equation}
and~$H^{(1)}_{\alpha}(\gamma)$ and~$H^{(1)'}_{\alpha}(\gamma) = \frac{1}{2}\left[ H^{(1)}_{\alpha-1}(\gamma) - H^{(1)}_{\alpha+1}(\gamma) \right]$ are Hankel functions of the first kind and their derivatives with respect to the argument, respectively.

Depending on the value of~$x$, both the order and the argument of the Hankel function can be considerably large.
While for small orders ($x\lesssim 100$) it is still feasible to use the exact numerical result, for higher orders the computation time becomes exceedingly long and we resort to analytical approximations.
First of all we introduce a new parameter~$\beta$ via~$e=\cos^{-1}\beta$. 
If~$x\tan^3\beta \gtrsim 1$ we use the standard approximation for Hankel functions with argument greater than the order
\begin{equation}
    H^{(1)}_{ix} \left(\frac{ix}{\cos\beta}\right) \simeq -i\left(\frac{2}{\pi x \tan\beta}\right)^{1/2} e^{-x(\tan\beta-\beta)} \sum_{n=0}^\infty \frac{\Gamma\left(n+\frac{1}{2}\right)}{\Gamma\left(\frac{1}{2}\right)} A_n(\beta) \left( \frac{-2}{x\tan\beta} \right)^n \, ,
\end{equation}
where we checked that a good accuracy can be achieved stopping at~$n=3$.
Coefficients in this approximation scheme read as
\begin{equation}
    \begin{aligned}
        A_0(\beta) &= 1 \, , \qquad A_1(\beta) = \frac{1}{8} + \frac{5}{24\tan^2\beta} \, , \\
        A_2(\beta) &= \frac{3}{128} + \frac{77}{576\tan^2\beta} + \frac{385}{3456\tan^4\beta} \, , \\
        A_3(\beta) &= \frac{5}{1024} + \frac{1521}{25600\tan^2\beta} + \frac{17017}{138240\tan^4\beta} + \frac{17017}{248832\tan^6\beta} \, .
    \end{aligned}
\end{equation}
Instead, if~$x \tan^3\beta \lesssim 1$ we use the approximation for when order and argument are nearly equal. 
More specifically, if~$x^{2/3}(1-\cos\beta)/(\cos\beta)^{2/3} \leq 1$ we use
\begin{equation}
    \begin{aligned}
        H^{(1)}_{ix} \left(\frac{ix}{\cos\beta}\right) &\simeq -\frac{2}{3\pi} \sum_{n=0}^\infty e^{\frac{2}{3}(n+1)\pi i} \sin\left(\frac{n+1}{3}\pi\right) \Gamma\left(\frac{n+1}{3}\right) \\
        &\qquad\qquad\qquad \times \left( \frac{6\cos\beta}{ix} \right)^{(n+1)/3} B_n\left( \frac{ix(1-\cos\beta)}{\cos\beta} \right) \, ,
    \end{aligned}
\end{equation}
where in this case we stop at~$n=5$ and the~$B_n$ coefficients read as
\begin{equation}
    \begin{aligned}
        B_0(y) &= 1 \, , \quad B_1(y) = y \, , \quad B_2(y) = \frac{y^2}{2}-\frac{1}{20} \, , \quad	B_3(y) = \frac{y^3}{6}-\frac{y}{15} \, ,	\\
        \quad B_4(y) &= \frac{y^4}{24}-\frac{y^2}{24}+\frac{1}{280} \, , \quad B_5(y) = \frac{y^5}{120}-\frac{y^3}{60}+\frac{43y}{8400} \, .
    \end{aligned}
\end{equation}
Otherwise we use the approximation
\begin{equation}
    H^{(1)}_{ix} \left(\frac{ix}{\cos\beta}\right) \simeq -i \frac{\tan\beta}{\sqrt{3}} e^{-x\left(\tan\beta-\beta-\frac{1}{3}\tan^3\beta\right) + \frac{2}{3}\pi i} H^{(1)}_{1/3} \left(-\frac{x\tan^3\beta}{3}\right) \, .
\end{equation}


\section{Probability distribution functions for hyperbolic encounters}
\label{app:hyperbolic_encounter_pdfs}

In this appendix we provide a semi-analytical estimate of~$p(r_\mathrm{ini}, \theta_\mathrm{ini}, v_\mathrm{rel})$, the probability distribution function required to model the hyperbolic encounters in section~\ref{subsec:hyperbolic_encounters}.
As a first approximation, we assume that the three stochastic variables are statistically independent, thus the total distribution factorizes as~$p(r_\mathrm{ini}, \theta_\mathrm{ini}, v_\mathrm{rel}) = p(r_\mathrm{ini}) p(\theta_\mathrm{ini}) p_\mathrm{tot}(v_\mathrm{rel})$.

Regarding the BH separation probability distribution function, instead of considering the ABH and PBH populations separately, we just consider a BH population with density
\begin{equation}
    n_\mathrm{tot}(r) \simeq 
    \left\lbrace \begin{aligned}
    &n_\mathrm{ABH} + n_\mathrm{PBH} \, , \qquad\qquad &r \leq r_\mathrm{0,PBH} \, , \\
    &n_\mathrm{ABH} \, , \qquad\qquad &r_\mathrm{0,PBH} < r \leq r_\mathrm{0,ABH} \, , \\
    &0 \qquad\qquad &r_\mathrm{0,ABH} < r \, .
\end{aligned} \right.
\end{equation}
Following the standard approach for an ideal gas of particles~\cite{Chandrasekhar:1943ws}, we derive the nearest neighbor distribution of BHs in the cluster core, i.e.,~the initial distance distribution, which reads as
\begin{equation}
    p(r_\mathrm{ini}) = \frac{3r^2_\mathrm{ini} \mathcal{N}_{r_\mathrm{ini}}}{a^3} e^{-(r_\mathrm{ini}/a)^3} \, ,
\end{equation}
where~$a = \left[ 3/(4\pi) (n_\mathrm{ABH} + n_\mathrm{PBH}) \right]^{1/3}$ is the Wigner-Seitz radius, and~$\mathcal{N}_{r_\mathrm{ini}}$ is a normalization constant obtained by restricting our region of interest to the inner part of the cluster core, i.e.,~allowing only~$r_\mathrm{ini} \in [r_\mathrm{min},r_\mathrm{max}]$. 
The minimum radius distance is defined as the one at which the dynamics of the two bodies is not dominated anymore by the cluster, i.e.,~when the potential energies satisfy~$G(M_1+M_2)/r_\mathrm{min} = \Psi_0$.
On the other hand, we arbitrarily define as maximum distance~$r_\mathrm{max} = \bar{r}/\alpha_\mathrm{cut-off}$, where~$\bar{r}=a \times \mathrm{Gamma}(4/3)$ is the average distance between BHs, $\mathrm{Gamma}(x)$ is the Gamma function and~$\alpha_\mathrm{cut-off}$ is a numerical factor that implicitly defines a cut-off radius.
We consider as threshold value~$\alpha_\mathrm{cut-off}=10$ to ensure that there is less than~$0.1\%$ probability of having a third BHs in the same volume.

The total relative velocity probability distribution function can be obtained by the applying Bayes theorem to the results derived in appendix~\ref{app:relative_velocity_pdf}.
Accounting for a mixed population of ABHs and PBHs with two different relative abundances and velocity dispersion, we find
\begin{equation}
    p_\mathrm{tot}(v_\mathrm{rel}) = \pi^2_\mathrm{ABH} \, p_\mathrm{ABH-ABH}(v_\mathrm{rel}) + 2 \pi_\mathrm{ABH} \, \pi_\mathrm{PBH} \, p_\mathrm{ABH-PBH}(v_\mathrm{rel}) + \pi^2_\mathrm{PBH} \, p_\mathrm{PBH-PBH}(v_\mathrm{rel}) \, ,
\end{equation}
where~$\pi_j=N_j/N_\mathrm{tot}=n_j/n_\mathrm{tot}$ are relative abundance probabilities.

Finally, since the vectors~$\boldsymbol{r}_\mathrm{ini}$, $\boldsymbol{v}_1$ and~$\boldsymbol{v}_2$ are expected to be uniformly distributed, it is straightforward to compute the initial angle~$\theta_\mathrm{ini}$ distribution.
However, we have to include the additional restriction that~$\cos\theta_\mathrm{ini}=\hat{\boldsymbol{v}}_\mathrm{rel} \cdot \hat{\boldsymbol{r}} < 0$, since in the opposite case the two bodies have already passed each other.
As for the case of the total relative velocity, the total initial angle distribution reads as
\begin{equation}
    p_\mathrm{tot}(\theta_\mathrm{ini}) = \pi^2_\mathrm{ABH} p_\mathrm{ABH-ABH}(\theta_\mathrm{ini}) + 2 \pi_\mathrm{ABH} \pi_\mathrm{PBH} p_\mathrm{ABH-PBH}(\cos\theta_\mathrm{ini}) + \pi^2_\mathrm{PBH} p_\mathrm{PBH-PBH}(\theta_\mathrm{ini}) \, .
\end{equation}

In order to ensure that the speed of the first BH remains subluminal we have to impose the constraint
\begin{equation}
    r_\mathrm{ini} \sin\theta_\mathrm{ini} \gtrsim \frac{(e_\mathrm{ini}+1)^{3/2}}{2(e_\mathrm{ini}-1)^{1/2}} r_S \, ,
\end{equation}
with~$r_S=2GM/c^2$ being the sum of the Schwarzschild radii of the two BHs, and~$e_\mathrm{ini}$ is the initial eccentricity given by
\begin{equation}
    e_\mathrm{ini} = \left[ 1 + \frac{v^4_\mathrm{rel} r^2_\mathrm{ini} \sin^2 \theta_\mathrm{ini}}{G^2M^2} \right]^{1/2} \, .
\end{equation}
Since~$r_S/r_\mathrm{ini} \ll 1$ is almost given by construction, we have that the bound is non-trivial only for values of the initial eccentricity very close to unity.
In that case, by using the explicit expression for the initial eccentricity, we obtain the bound
\begin{equation}
    \sin\theta_\mathrm{ini} \gtrsim \frac{r_S/r_\mathrm{ini}}{v_\mathrm{rel}/c} \, .
\end{equation}

Moreover, we have to ensure that the system remains unbound.
The energy loss due to GWs is
\begin{equation}
    \Delta E = - \frac{8\mu^2v^7_\mathrm{rel}}{15 M c^5} g(e_\mathrm{ini}) \, ,
\end{equation}
where
\begin{equation}
    g(e_\mathrm{ini}) = \frac{24\arccos\left(-\frac{1}{e_\mathrm{ini}}\right) \left(1 + \frac{73}{24}e^2_\mathrm{ini} + \frac{37}{96}e^4_\mathrm{ini} \right) + \sqrt{e^2_\mathrm{ini}-1} \left(\frac{301}{6} + \frac{673}{12}e^2_\mathrm{ini} \right)}{(e^2_\mathrm{ini}-1)^{7/2}} \, .
\end{equation}
Since the energy of the system at initial time is~$E_\mathrm{ini}=\frac{1}{2}\mu v^2_\mathrm{rel}$, the condition to impose to have~$E_\mathrm{fin}>0$ is
\begin{equation}
    1 - \frac{16}{15} \frac{\mu}{M} \left( \frac{v_\mathrm{rel}}{c} \right)^5 g(e_\mathrm{ini}) \gtrsim 0 \, . 
\end{equation}
Away from the~$e_\mathrm{ini}\to 1$ limit, the condition is very easily satisfied because of the relative velocity suppression.
However, when the eccentricity value gets close to unity, we find the extra condition
\begin{equation}
    \sin\theta_\mathrm{ini} \gtrsim \left( \frac{340\pi}{384} \right)^{1/7} \left( \frac{\mu}{M} \right)^{1/7} \frac{r_S/r_\mathrm{ini}}{(v_\mathrm{rel}/c)^{9/7}} \, ,
\end{equation}
which turns out to be stronger than the previous condition, i.e.,~if this condition is satisfied, the previous one is also satisfied.

\begin{figure}[ht]
    \centerline{
    \includegraphics[width=\columnwidth]{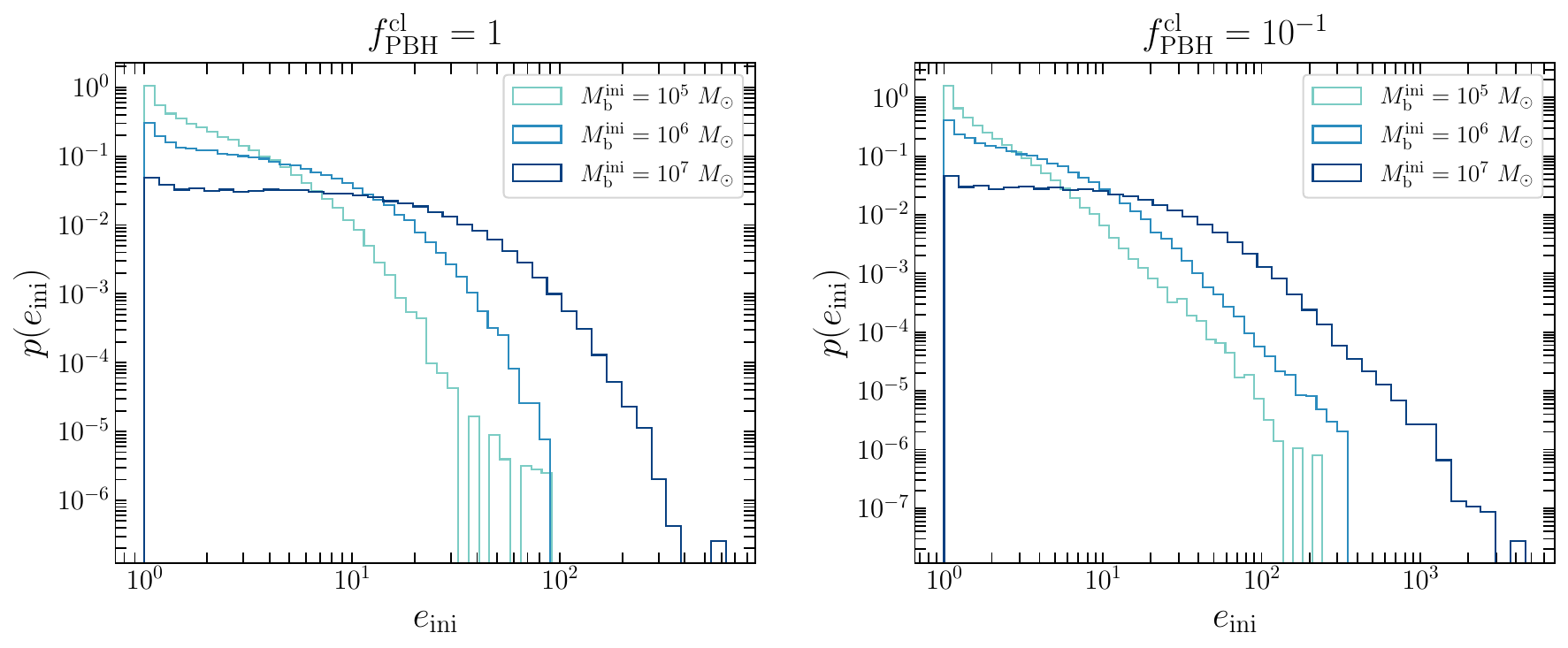}}
    \caption{Initial eccentricity distribution for GCs with~$M_\mathrm{PBH}=100 \,  M_\odot$ and different PBH abundance, i.e.,~$f^\mathrm{cl}_\mathrm{PBH}=1$ (\textit{left panel}) and~$f^\mathrm{cl}_\mathrm{PBH}=10^{-1}$ (\textit{right panel}).}
    \label{fig:eccentricity_distribution}
\end{figure}

Finally, we show in figure~\ref{fig:eccentricity_distribution} the distribution of initial eccentricity for GCs with different masses and PBH abundance.
Also in this case, the eccentricity probability distribution function is computed from a catalog of hyperbolic encounters with parameters sampled from the distribution discussed above.
The emission of GW due to hyperbolic encounters is peaked at values of eccentricity close to unity; however, as we can see from the figure, such values are probable only in low-mass GCs.
Therefore, it is reasonable to expect that the overall GWB produced by this class of dynamical interactions in this environment will not be comparable to other, more effective, emission channels.


\bibliography{biblio}
\bibliographystyle{utcaps}


\end{document}